\documentclass[12pt]{amsart}

\usepackage[margin=2.5cm]{geometry}
\usepackage{tikz}
\tikzset{node distance=2cm, auto}
\usetikzlibrary{matrix,arrows,decorations.pathmorphing}

\usepackage{graphicx}
\usepackage{amsmath,amssymb}

\usepackage{amscd}
\usepackage{verbatim}
\usepackage{enumerate}

\newtheorem{conjecture}{Conjecture}

\newcommand{\RR}{\mathbb{R}}

\newcommand{\ZZ}{\mathbb{Z}}

\newcommand{\cF}{{\mathcal F}}

\newcommand{\pa}{\partial}

\newcommand{\be}{\begin{equation}}
\newcommand{\ee}{\end{equation}}

\newcommand{\cL}{\mathcal{L}}

\newcommand{\cP}{\mathcal P}

\newenvironment{dedication}
        {\vspace{6ex}\begin{quotation}\begin{center}\begin{em}}
        {\par\end{em}\end{center}\end{quotation}}

\title{Poisson sigma model and semiclassical quantization of integrable systems.}

\author{Alberto S. Cattaneo}
\address{A.S.C.: Institut for Mathematik, Winterthurerstrasse 190, 8057 Zurich, Switzerland}
\email{cattaneo@math.uzh.ch}

\author{Pavel Mnev}\address{P.M.: University of Notre Dame, Notre Dame, Indiana 46556, USA \&
St. Petersburg Department of V. A. Steklov Institute of Mathematics of the Russian
Academy of Sciences, Fontanka 27, St. Petersburg, 191023 Russia}
\email{ pmnev @nd.edu}

\author{Nicolai Reshetikhin}
\address{N.R.: Department of Mathematics, University of California, Berkeley,
CA 94720, USA \& KdV Institute for Mathematics, University of Amsterdam,
Science Park 904, 1098 XH Amsterdam, The Netherlands \& QGM, Aarhus University, Denmark \& Physics Department, St. Petersburg University, Russia }
\email{reshetik@math.berkeley.edu}

\begin{document}

\maketitle

\begin{dedication}
\hspace{4cm}
\vspace*{3cm}{Dedicated to the memory of L.D. Faddeev.}
\end{dedication}

\begin{abstract}
In this paper we outline the construction of semiclassical eigenfunctions
of integrable models in terms of the semiclassical path integral for
the Poisson sigma model with the target space being the phase space of the integrable system.
The semiclassical path integral is defined as a formal power series with coefficients being Feynman diagrams.
We also argue that in a similar way one can obtain irreducible semiclassical representations of Kontsevich's star
product.
\end{abstract}

\section*{Introduction} This paper is motivated by two problems.
The first one is how to describe
the full semiclassical expansion of eigenfunctions of a complete set of quantum Hamiltonians for
a quantum integrable system. More generally, the problem is how to describe full semiclassical
asymptotics of the scalar products of two integrable systems on the same space of geometric quantization,
or corresponding cyclic amplitudes.
The second problem is how to construct an irreducible semiclassic representation of Kontsevich's
star product on the algebra of functions on a Poisson manifold.

It turns out that both questions can be answered in terms the semiclassical quantization of
the Poisson sigma model with special boundary conditions. This connects the structure of the semiclassical eigenfunctions of any given
integrable system to the quantization of a very special topological gauge theory.

This paper is a research announcement. An extended version of this paper with proofs and mathematically sound construction
of Feynman diagrams will be published separately.

We would like to dedicate this paper to the memory of L.D. Faddeev. He made fundamental contributions to both, quantization of gauge theories and
quantization of integrable systems \cite{FP}\cite{F}. As we demonstrate here they are related in a very direct way.
For other relations between gauge theories and integrable systems see \cite{CMW}\cite{Lip}\cite{Ka}\cite{NS}.

{\bf Acknowledgements}  A. S. C. acknowledges partial support of SNF Grant No. 200020-172498/1.
This research was (partly) supported by the NCCR SwissMAP, funded by the Swiss National Science Foundation, and by the COST Action MP1405 QSPACE, supported by COST (European Cooperation in Science and Technology).  P. M. acknowledges partial support of RFBR Grant No. 17-01-00283a. The work of N.R. was supported by the NSF grant DMS-1601947.

\section{Semiclassical asymptotic in integrable systems and topological quantum mechanics}

\subsection{ Semiclassical asymptotic in integrable systems}

 Let $(M, \omega)$ be a symplectic manifold. Fix the geometric quantization data
which consist of the following:

\begin{itemize}
\item A line bundle $L$ (a prequantization line bundle) with Hermitian structure
on fibers and a Hermitian connection $\alpha$ on $L$ such that the symplectic
form $\omega$ is the curvature of $\alpha$, i.e. $d\alpha=\omega$ and the Hermitian
product is covariantly constant.

\item A real polarization $P\subset TM$ which is an integrable
tangent distribution on $M$ such that each generic leaf is a Lagrangian submanifold
in $M$. We assume that the space of leaves $B=M/P$ is almost everywhere smooth,
i.e. that the polarization is a Lagrangian fibration $\pi: M\to B$ (generic fibers are Lagrangian).

\end{itemize}

The space of geometric quantization $H^{(1/2)}_P$ is
the space of half-densities on $M$ which are covariantly constant (with respect to the connection $\alpha$)
along $P$. Locally it can be identified with functions on $M/P$.

Let $C_h(M)$ be quantized algebra of functions on $M$. Assume it acts on the space $H^{(1/2)}_P$.

A classical integrable system on $M$ is a Lagrangian fibration\footnote{The generic fibers are Lagrangian
submanifolds.} $\pi: M\to B$ which defines a Poisson
commuting  subalgebra $C(M,B)\subset C(M)$ in the algebra of functions on $M$,  $C(M,B)=\pi^*(C(B))$.
Assume that the subalgebra $C(M,B)$ is quantized, i.e. deformed into a maximal commutative subalgebra $C_h(M,B)\subset C_h(M)$, where $C_h(M)$ is
a deformation quantization of $C(M)$ given by a star product (see \cite{K} and references therein). For $2n$-dimensional $M$ such subalgebra has rank $n$. For the purpose of this paper one can think 
that $M=T^*Q$ for some smooth $n$-dimensional manifold $Q$, the algebra $C_h(M)$ is the algebra
of differential operators where derivatives are multiplied by $-ih$ and the algebra $C(M)$ is the algebra of 
smooth functions on $Q$ which are polynomials in the cotangent directions. If $h$ is a formal variable we assume that vector
spaces $C_h(M)$ and $C(M)[[h]]$ are isomorphic and the isomorphism is given  by placing derivatives to the right of the 
smooth coefficient functions. The commutative subalgebra $C_h(M,B)$
is the algebra of commuting differential operators with the principal symbol from $\pi^*(C(B))$.

Now, assume that we have a real polarization $P$ with the corresponding Lagrangian fibration $\pi: M\to B=M/P$. Assume we have two integrable systems corresponding to the Lagrangian fibrations $\pi_{1,2}: M\to B_{1,2}$\footnote{ In other words we have three mutually transverse polarizations.}. Assume that fibers of both projections are generically transverse and generically transverse with leaves of $P$. Assume that the algebra $C_h(M)$ acts on the space $H^{(1/2)}_P$. When $M=T^*Q$, $P$ is the real polarization given by the Lagrangian fibration $T^*Q\to Q$
and the space $H^{(1/2)}_P$ can be naturally identified with half-densities on $Q$.

We will say a vector $\psi_\chi\in H^{(1/2)}_P$ is an eigenvector of $C_h(M,B_2)$
corresponding to the character $\chi: C_h(M,B_2)\to \RR$ if
$a\psi_\chi=\chi(a)\psi_\chi$ for any $a\in C_h(M,B_2)$. Semiclassically, as $h\to 0$ the set of
characters can be identified with $B_2$. Similarly, we have eigen-half-densities for $C_h(M,B_1)$.
Semiclassically, characters of $C_h(M,B_1)$ can be identified with points of $B_1$. Denote by $\psi^{(1)}_{b_1}$
and $\psi^{(2)}_{b_2}$ semiclassical eigen-half-densities of commutative subalgebras $C_h(M,B_1)$ and $C_h(M,B_2)$
acting in $H^{(1/2)}_P$.

\begin{conjecture}\cite{R} Semiclassical asymptotics of the scalar product of eigen-half-densities of $C_h(M,B_2)$
and $C_h(M,B_1)$ in the space $H^{(1/2)}_P$
has the following structure:

\begin{eqnarray}\label{sc-product}
(\psi^{(2)}_{b_2}, \psi^{(1)}_{b_1})=\frac{C}{(2\pi h)^{n/2}}&\sum_{c\in \cL_{b_1}^{(1)}\cap \cL_{b_2}^{(2)}} e^{\frac{i}{h}S_{\gamma_1,\gamma_2}(c,b_1,b_2)+\frac{i\pi}{2}\mu_{\gamma_1,\gamma_2}(c)}\\ &\left|det\left(\frac{\pa^2 S_{\gamma_1,\gamma_2}(c,b_1,b_2)}{\pa b_1^i\pa b_2^j}\right)\right|^{1/2}(1+O(h))\sqrt{|db_1db_2|} \nonumber
\end{eqnarray}

Here as we made a choice of reference points $x_0^{(i)}\in \cL_{b_i}^{(i)}$,  $S_{\gamma_1,\gamma_2}(c,b_1,b_2)=S^{(1)}_{\gamma_1}-S^{(2)}_{\gamma_2}$ where $S^{(i)}_{\gamma_i}=\int_{\gamma_i\subset \cL^{(i)}}\alpha$ and $\alpha$ is the prequantization connection, $\gamma_1$ and $\gamma_2$ are
paths in $\cL^{(1)}_{b_1}$ and $\cL^{(2)}_{b_2}$ respectively, connecting corresponding reference points\footnote{Here we assume that paths are contained in a neighborhood where the bundle
is trivialized and $\alpha$ is a $1$-form on $M$. Otherwise we should lift paths to the prequantization bundle.  The result gives the parallel transport in the prequantization line bundle which is
consistent with the fact that globally, $\psi^{(a)}_{b_a}$ are not functions in $q$ abd $b_a$, but sections
of the prequantization line bundle. Globally, the exponent in the formula (\ref{sc-product}) is also not a function in
$b_1$ and $b_2$, but a section
of a line bundle.

Changing $\alpha\mapsto \alpha+ df$ is an automorphism of the prequantization line bundle. It corresponds to the shift $-ih\frac{\pa}{\pa q}\mapsto ih\frac{\pa}{\pa q}+\frac{\pa f}{\pa q}$ in the deformation quantization of $C(T^*Q)$ by differential operators. In (\ref{sc-product}) it results in the phase change $(\psi^{(2)}_{b_2}, \psi^{(1)}_{b_1})\mapsto (\psi^{(2)}_{b_2}, \psi^{(1)}_{b_1})\exp(\frac{i}{h}(f(x_1)-f(x_2))$. } and $c$.
Lagrangian submanifolds $\cL^{(1)}_{b_1}$ and  $\cL^{(2)}_{b_2}$ are fibers over $b_1\in B_1$ and $b_2\in B_2$ respectively. We assume that $\cL_{b_1}^{(1)}$ and $\cL_{b_2}^{(2)}$ are transverse. The number $\mu_{\gamma_1,\gamma_2}(c)$
is the weighted number of points along $\gamma_2$ between $x_2$ and $c$ where the tangent space to $\cL^{(2)}_{b_2}$ intersect a fiber of $\pi_1$ over a line. The point counts with plus if it is crossed
in the positive direction and with the minus if it is crossed in the negative direction, i.e. $\mu$ is the Maslov index.
When Bohr-Sommerfeld quantization conditions for $b_1$ and $b_2$ hold, the exponent does not depend
on the choice of $\gamma_1$ and $\gamma_2$. The Hessian and the higher order terms do not depend on the choice of reference points.

\end{conjecture}

Note that to write the formula (\ref{sc-product}) we made a choice of a reference point
on every fiber of $\pi_1$ and of $\pi_2$. Consistent choice of such reference points
can be inferred, e.g., from fixing a Lagrangian submanifold $\Lambda$ intersecting
fibers of $\pi_1$ and $\pi_2$ generically transversally. A change of $\Lambda$ changes the formula for the scalar product by a constant factor (each term in the sum changes by the same factor). Thus, the eigenfunctions depend on $\Lambda$
only projectively. The choice of $\Lambda$ is also convenient for making the expression (\ref{sc-product}) invariant
with respect to automorphisms of the prequantization line bundle. Let us multiply the scalar product by the factor $\exp(\frac{i}{h}\int_{\gamma\in\Lambda}\alpha)$ where the path $\gamma$ connects intersection points $\Lambda\cap \cL_1$ and $\Lambda\cap \cL_2$ in $\Lambda$. From the spectral point of view this is just the multiplication of eigenfunctions by constants.
However, it makes semiclassical scalar products invariant with respect to automorphisms of the prequantization line bundle.

A natural object which does not depend on the choice of $\Lambda$ is the transition probability, i.e. the square of the absolute value of the scalar product:
\begin{eqnarray}\label{prob}
|(\psi^{(2)}_{b_2}, \psi^{(1)}_{b_1})|^2=\frac{1}{{2\pi h}^n}\sum_{a,c\in \cL_{b_1}^{(1)}\cap \cL_{b_2}^{(2)}}
\left|det\left(\frac{\pa^2 S_\gamma(c,b_1,b_2)}{\pa b_1^i\pa b_2^j}\right)det\left(\frac{\pa^2 S_\gamma(a,b_1,b_2)}{\pa b_1^i\pa b_2^j}\right)\right|^{1/2}\\
\exp\left(\frac{i}{h}S_{\gamma_1,\gamma_2}(b_1,b_2)+
\frac{i\pi}{2}(\mu_{\gamma_1,\gamma_2}(c)-
\mu_{\gamma_1,\gamma_2}(a))\right)(1+O(h))
|db_1db_2| \nonumber
\end{eqnarray}
Here $\gamma_1$ is a path in $\cL_{b_1}^{(1)}$ connecting points $a$ and $c$ and $\gamma_2$ is a path in $\cL_{b_2}^{(2)}$ connecting these points. The exponent $S_{\gamma_1,\gamma_2}(b_1,b_2)=\int_{D_{a,c}}\omega$, where $D_{a,c}$ is a disc bounded by $\gamma_1$ and $\gamma_2$, does not depend on
the choice of paths if Bohr-Sommerfeld quantization conditions hold. The Hessian also does not depend on the choice of $\gamma$'s.

 \begin{figure}[htb]
\includegraphics[height=5cm,width=7cm]{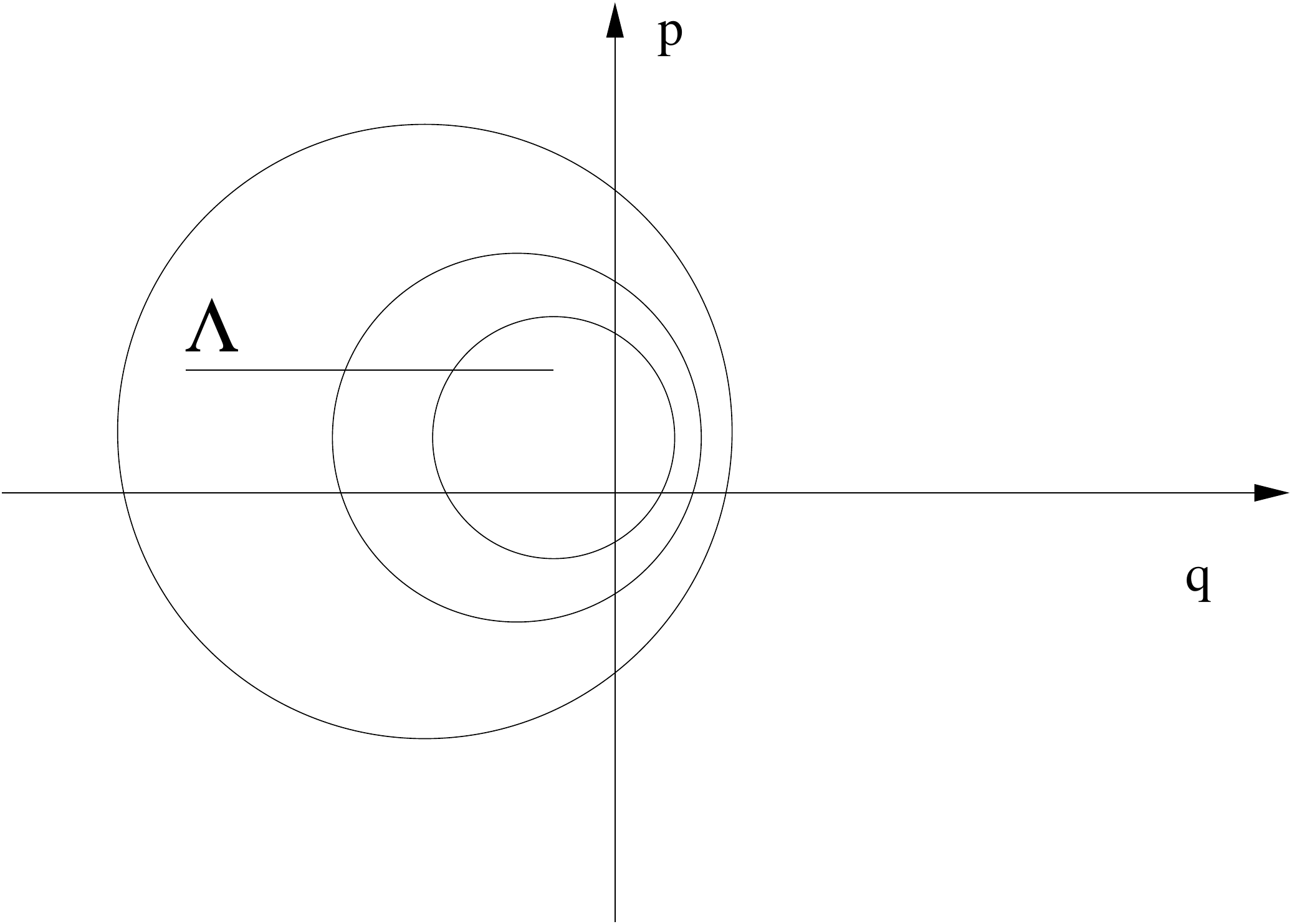}
\caption{Choice of reference points corresponding to an integrable system
on $\RR^2=T^*\RR$ with $\alpha=pdq$ and the standard Lagrangian fibration $T^*\RR\to \RR$. In this
case $\Lambda=\{p=0\}$. Locally, $\Lambda$ is a section of the Lagrangian
fibration given by  level curves of the Hamiltonian. }
\label{F-1}
\end{figure}

More generally, if $\pi_a: M\to B_a$ are Lagrangian fibrations, the cyclic product
\[
(\psi^{(1)}_{b_1}, \psi^{(k)}_{b_{k}})\dots (\psi^{(3)}_{b_3}, \psi^{(2)}_{b_2}) (\psi^{(2)}_{b_2}, \psi^{(1)}_{b_1})
\]
does not depend on the choice of $U(1)$-ambiguities of normalized eigenfunctions. We will call such products
{\it cyclic amplitudes}. For the semiclassical asymptotic of a cyclic amplitude we have:

\begin{eqnarray}\label{cycl-amp}
&(\psi^{(1)}_{b_1}, \psi^{(k)}_{b_{k}}) \dots (\psi^{(3)}_{b_3}, \psi^{(2)}_{b_2}) (\psi^{(2)}_{b_2}, \psi^{(1)}_{b_1})=\\ \nonumber
&\frac{1}{{2\pi h}^{kn/2}}\sum_{c_1\in \cL_{b_1}^{(1)}\cap \cL_{b_2}^{(2)}, \dots, c_k\in \cL_{b_k}^{(k)}\cap \cL_{b_1}^{(1)}}
\left| \prod_{a=1}^k det\left(\frac{\pa^2 S}{\pa b_a^i\pa b_{a+1}^j}\right)\right|^{1/2} \\
&\exp\left(\frac{i}{h}S(c_1,\dots c_k)+
\frac{i\pi}{2}\mu\right)(1+O(h))
\prod_{a=1}^k |db_a| \nonumber
\end{eqnarray}
Here $S(c_1,\dots, c_k)=\sum_{a=1}^k\int_{\gamma_{a,a+1}}\alpha|_{\cL^{(a)}_{b_a}}$ where $\gamma_{a,a+1}$ is a path in
$\cL^{(a)}_{b_a}$ connecting $c_a$ and $c_{a+1}$ and $\mu$ is the corresponding Maslov index.
The exponent does not depend on the choice of $\gamma$'s
if Bohr-Sommerfeld conditions hold. By the Stokes theorem it is clear that  $S(c_1,\dots, c_k)=\int_D \omega$ where
$D$ is a disk bounded by paths $\gamma_{a,a+1}$.

\subsection{Topological quantum mechanics}
\subsubsection{Scalar product as a path integral for topological quantum mechanics}
We want to write the formula  (\ref{sc-product}) as the semiclassical expansion of a path integral. The natural candidate is
the topological quantum mechanics
\begin{equation}\label{PI-TQM}
(\psi^{(2)}_{b_2}, \psi^{(1)}_{b_1})=C\int_{\gamma(0)\in \cL_1, \gamma(1)\in \cL_2} e^{\frac{i}{h}(\int_\gamma \alpha +f^{(1)}_{b_1}(\gamma(0))-f^{(2)}_{b_2}(\gamma(1))+\int_{\gamma_0\subset \Lambda}\alpha )} D\gamma
\end{equation}
Here $\gamma_0$ connects reference points $x_1$ and $x_2$\footnote{The last summand in the exponent is making it to be invariant with respect to automorphisms of the prequantization line bundle. This means that the integral and 
the expression (\ref{sc-product}) with this factor depend only on the curvature $\omega=d\alpha$ of $\alpha$.} and $\cL_a=\cL^{(a)}_{b_a}$ where $a=1,2$ and $f^{a}_{b_a}$ are boundary corrections with defining property
\begin{equation}\label{f-eq}
df^{(a)}_{b_a}=\iota^*_a(\alpha)
\end{equation}
here $\iota_a: \cL_a \to M$ are embeddings of Lagrangian submanifolds to $M$. These functions can be chosen as
follows. For $x_a\in \cL_a$ fix a path $\gamma_a$ on $\cL_a$ connecting $x$ with a reference point $x_a\in \cL_a$.
Define
\[
f^{(a)}_{b_a}(x)=\int_{\gamma_a}\alpha
\]
This function satisfies (\ref{f-eq}) and the exponent in (\ref{PI-TQM}) does not depend on $\gamma_a$
if Bohr-Sommerfeld quantization conditions hold.
An arbitrary constant $C$ changes by the factor $exp(\frac{i}{h}(\int_{x_1}^{y_1}\alpha-\int_{x_2}^{y_2}\alpha))$
when we change reference points from $x$ to $y$. 

This integral should be understood semiclassically, as a formal power series supported at
critical points of the exponent.  Changing the reference points $x_a$ changes the integral by a constant factor.
A consistent choice of reference points on fibers of Lagrangian fibrations $\pi_a: M\to B_a$ is given a Lagrangian submanifold $\Lambda$ intersecting transversally each fiber.

It is easy to see that critical points are exactly the intersection points $\cL_1\cap \cL_2$. Indeed if we choose
local coordinates $x^i$ so that $\alpha=\sum_i \alpha_i(x)dx^i$ and choose a parametrization of $\gamma$, we will have
the following formula for the variation of the integral in the exponent:
\[
\delta \int_\gamma\alpha= \int_0^1\sum_{ij}(\frac{\alpha_i}{\pa x^j}-\frac{\alpha_j}{\pa x^i})\delta x^j(t) \frac{dx^i}{dt}(t) dt+ \sum_i(\alpha_i(x(1))\delta x^i(1)-\alpha_i(x(0))\delta x^i(0))
\]
This implies that $\frac{dx^i}{dt}=0$, i.e. critical points are constant trajectories and therefore, taking into account boundary conditions for $\gamma$, we conclude that  the critical points of the exponent in (\ref{PI-TQM})
are the intersection points $\cL_1\cap \cL_2$. However, to develop the perturbation theory for such path integrals it is convenient to
reformulate them in terms of two dimensional topological field theory. So, the next step is to reformulate the integral (\ref{PI-TQM}) as the partition function for the Poisson sigma model with special boundary conditions. We will do this
 in the section \ref{PSM}.

\subsubsection{Transition probabilities and cyclic amplitudes}

Transition probabilities can also be written in terms of topological quantum mechanics.
They do not depend on the normalization of eigenfunctions. Semiclassically this leads to an unambiguous
semiclassical formula (\ref{prob}) for these probabilities which does not require a choice of references points.
It is natural to expect the following path integral for (\ref{prob}):
\begin{equation}\label{d-tqm}
|(\psi^{(2)}_{b_2}, \psi^{(1)}_{b_1})|^2=\int_{\gamma_1(0),\gamma_2(0)\in \cL_1; \gamma_1(1),\gamma_2(1)\in \cL_2} \exp\left(\frac{i}{h}\left(\int_{\gamma_1} \alpha -\int_{\gamma_2}\alpha+\int_{\sigma_1}\alpha-\int_{\sigma_2}\alpha\right)\right) D\gamma D\sigma
\end{equation}
Here path $\sigma_1$ connecting $\gamma_1(0)$ and $\gamma_2(0)$ in $\cL_1$ and $\sigma_2$ connecting $\gamma_1(1)$ and $\gamma_2(1)$ in $\cL_2$.
As for the amplitudes, this path integral should be understdood semiclassically. We argue that
it reproduces the formula (\ref{prob}).

Similarly we expect that the cyclic amplitudes can be written as the path integral
\begin{eqnarray}\label{cycl-tqm}
&(\psi^{(1)}_{b_1}, \psi^{(k)}_{b_{k}}) \dots (\psi^{(3)}_{b_3}, \psi^{(2)}_{b_2}) (\psi^{(2)}_{b_2}, \psi^{(1)}_{b_1})=\\
&\int \exp\left(\frac{i}{h}\sum_{a=1}^k \left(\int_{\gamma_a} \alpha +\int_{\sigma_a}\alpha\right)\right) \prod_{a+1}^kD\gamma_a D\sigma_a \nonumber
\end{eqnarray}
Here $\gamma_a: I=[0,1]\to M$, such that $\gamma_a(0)\in \cL_a$, $\gamma_a(1)\in \cL_{a+1}$ and $\sigma_a$ is a path in $\cL_a$ such that
$\sigma_a(0)=\gamma_{a-1}(1), \ \ \sigma_a(1)=\gamma_{a}(0)$.

\subsubsection{Two-dimensional version of the path integral for the topological quantum mechanics}
Here we reformulate  the topological classical mechanics as a two-dimensional topological field theory.
As before, $M$ is a symplectic manifold with a prequantization line bundle with connection $\alpha$ such that $\omega=d\alpha$.
Consider the classical field
theory on a disk $D$ with fields $X: D\to M$ and with the action functional\footnote{ This action is also the topological part of the A-model \cite{W}.}
\[
S[X]=\frac{1}{2}\int_DX^*(\omega)=\frac{1}{2}\int_D \omega_{ij}(X)dX^i\wedge dX^j
\]
Using Stokes theorem we can write it as:
\[
\int_DX^*(d\alpha)=\int_{\pa D}X^*(\alpha),
\]
where $\alpha$ is the prequantization connection.
Now let us organize the boundary conditions in such a way that the integral over the boundary $\pa D$
would give the exponent in (\ref{PI-TQM}).  We assume that $\Lambda$ is a Lagrangian submanifold defining the  reference
points. It should be transversal to
$\cL_{b_1}$ and $\cL_{b_2}$ for generic $(b_1,b_2)$.

Assume the boundary is partitioned in four intervals $\pa D=I\sqcup I_1\sqcup I_2\sqcup I_{12}$ with no restriction on $X$ on $I$ and
\[
X|_{I_1}\in \cL_1, \ \ X|_{I_2}\in \cL_2, \ \ X|_{I_{12}}\in \Lambda
\]
\begin{figure}[htb]
\includegraphics[height=6cm,width=8cm]{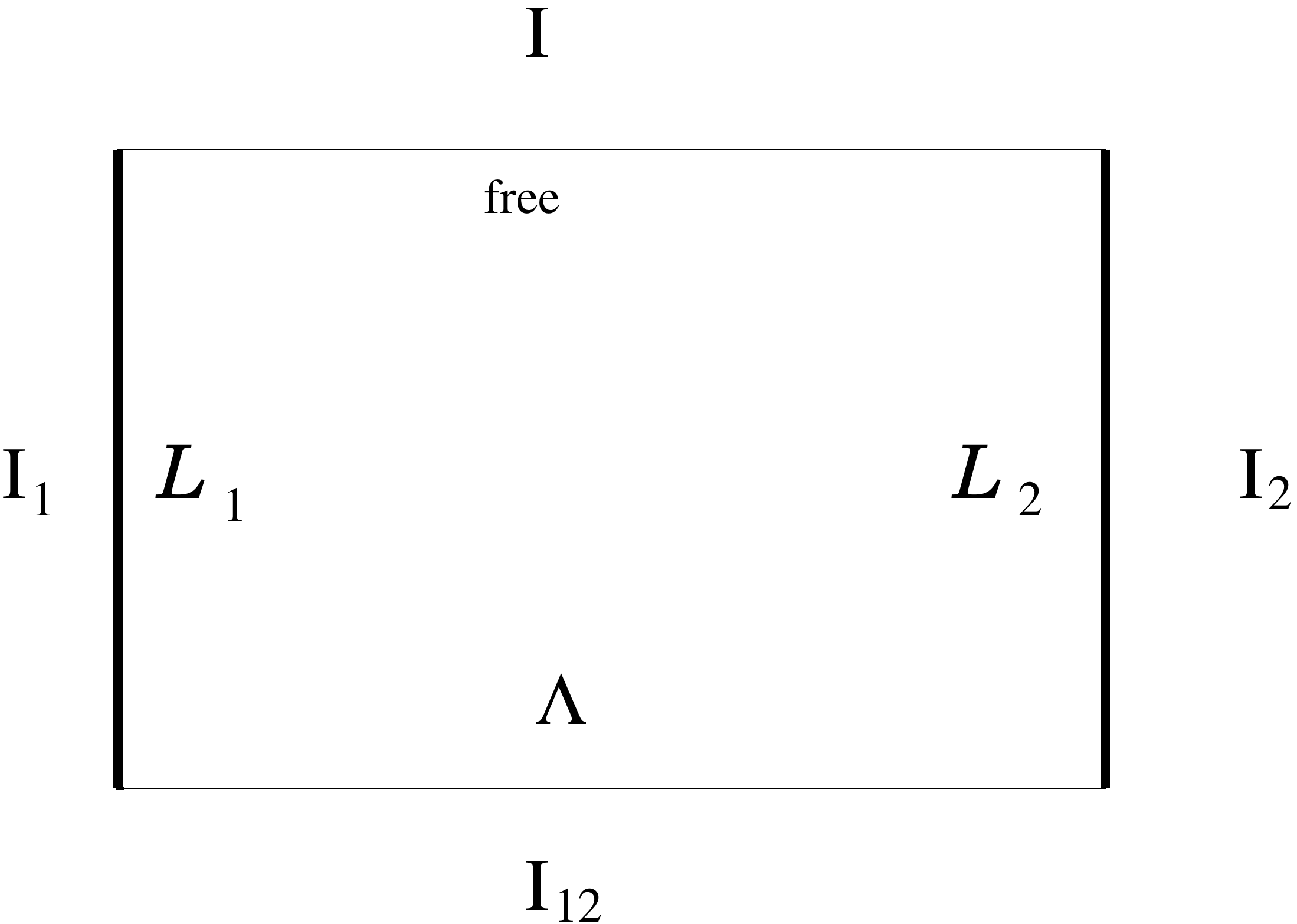}
\caption{Partitioning the boundary of a disk and the boundary conditions for the field $X$ corresponding to amplitudes in topological quantum mechanics.}
\label{F-2}
\end{figure}
Then the integral over the boundary can be written as
\[
\int_{\pa D}\alpha=\int_I\alpha+\int_{I_1}\alpha+\int_{I_2}\alpha+\int_{I_{12}}\alpha=\int_I\alpha+f_{b_1}^{(1)}(\gamma(0))-f_{b_2}^{(2)}(\gamma(1))+\int_{I_{12}}\alpha
\]
where $f_1$ and $f_2$ are generating functions for the connection $\alpha$ restricted to the Lagrangian
submanifolds $\cL_1$ and $\cL_2$ respectively as in (\ref{PI-TQM}), with reference points $x_i=\cL_i\cap \Lambda$.
The last term corresponds to an arbitrary constant $C$ in (\ref{f-eq}). Denote by $\gamma=\{X(s)\in M|s\in I\}$ the parameterized path in $M$ corresponding to $I\subset \pa D$, then we can write
\[
\int_IX^*(\alpha)=\int_{\gamma}\alpha, \qquad f^{(i)}_{b_i}(x)=\int_{\gamma_{x,x_i}}\alpha,
\]
Here $x\in \cL_i$, $x_i=\cL_i\cap \Lambda$.

Critical points of $S[X]$ are mappings which are constant on $I$ and satisfy the boundary conditions on the other
parts of $\pa D$.

The path integral representing the scalar product (\ref{PI-TQM})
can be written then up to (an infinite) constant as
\begin{equation}\label{A}
(\psi^{(2)}_{b_2}, \psi^{(1)}_{b_1})=\int_{X:D\to M} e^{\frac{i}{h}\int_D X^*(\omega)} DX
\end{equation}
with the boundary conditions described above.

This path integral should be treated as a semiclassical path integral
given by oscillatory exponential contributions and power series with coefficients being Feynman diagrams.
It is a second order topological
quantum field theory. To deal with the boundary conditions using the Hamiltonian framework, we will reformulate it as the first order Poisson sigma model.

Note that we expect to reproduce the formula (\ref{sc-product}) where we initially fixed the
reference points $x_a\in \cL_{b_a}\cap \Lambda$. It means that we should fix  $X(u_a)=x_a$,
where $u_a=I_a\cap I_{12}$, in the formula (\ref{A}). The path integral then is defined
in terms of Feynman diagram contributions for each critical point of the action with fixed
$x_1$ and $x_2$. Varying $x_1$ and $x_2$ results in an overall $U(1)$ factor and does
not change the Feynamn diagram contributions.

\begin{figure}[htb]
\includegraphics[height=7cm,width=7cm]{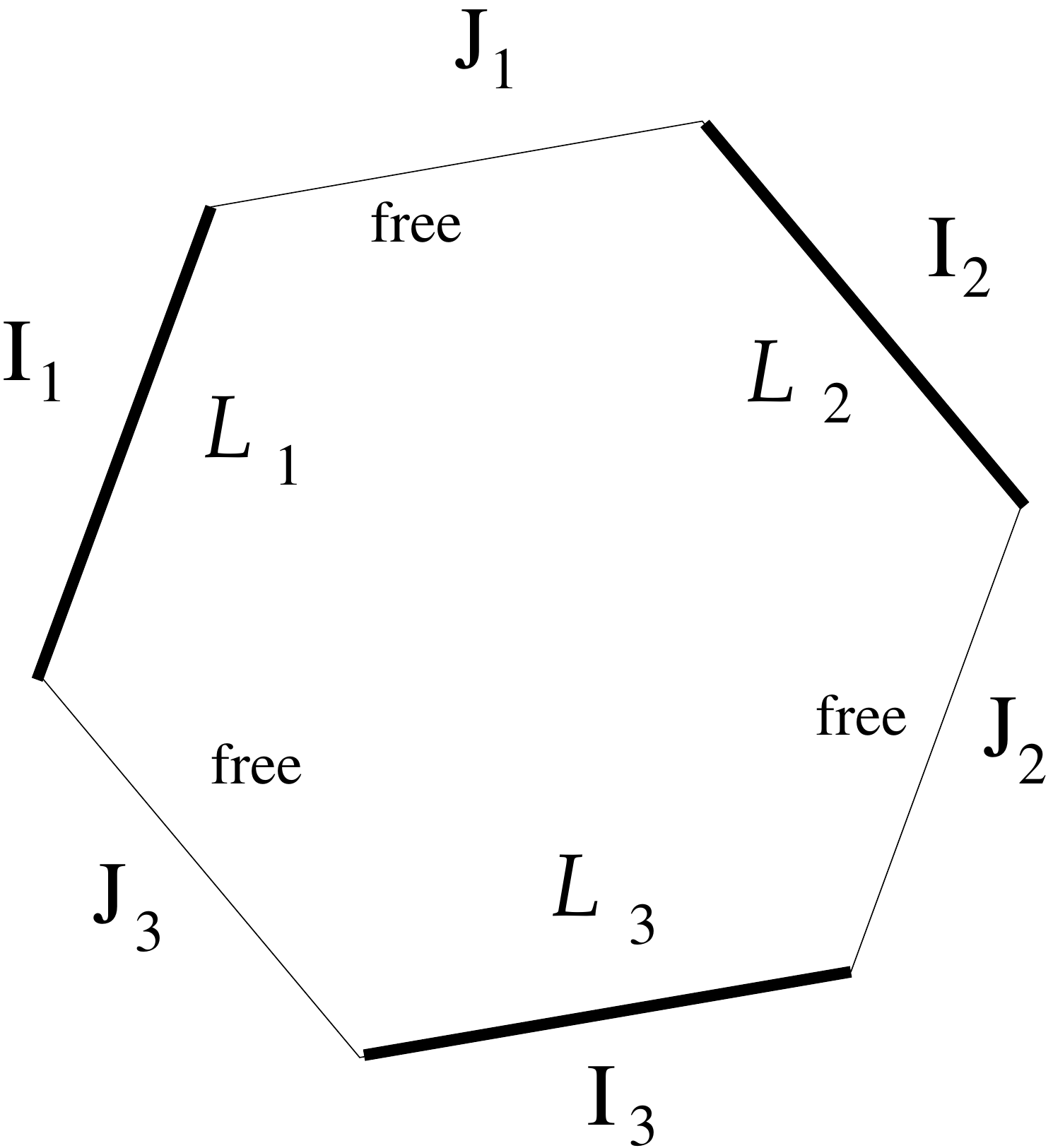}
\caption{A partitioning of the boundary of a disk and the boundary conditions for the $X$-field corresponding to cyclic amplitudes, $k=3$.}
\label{cyclic}
\end{figure}

Similarly the path integral for the topological quantum mechanics describing the cyclic amplitudes
can be written as

\begin{equation}\label{cycl-tqm}
(\psi^{(1)}_{b_1}, \psi^{(k)}_{b_{k}}) \dots (\psi^{(3)}_{b_3}, \psi^{(2)}_{b_2}) (\psi^{(2)}_{b_2}, \psi^{(1)}_{b_1})=
\int_{X:D\to M} e^{\frac{i}{h}\int_D \omega} DX
\end{equation}
where the boundary of the disc is partitioned to intervals $I_1,\dots I_k$ and $J_1, \dots, J_k$, see Fig.\ref{cyclic} with boundary
conditions $X|_{I_a}\in \cL_a$ and $X$ is free on intervals $J_a$.

\section{Topological quantum mechanics as the Poisson sigma model}\label{PSM}

\subsection{Classical Poisson sigma model corresponding to topological quantum mechanics}

\subsubsection{Poisson sigma model and boundary conditions}\label{PSM-bc} Recall that in the Poisson sigma model, the fields are pairs $(X,\eta)$ representing components of
the mapping of bundles $(X,\eta): TD\to T^*M$. Because the main difficulty of the problem
is not in the geometric structure of $M$, we assume that $M=\RR^N$ for some even $N$, equipped with possibly nonconstant symplectic structure. Then $X:D\to \RR^N$ and
$\eta\in \Omega^1(D)\otimes \RR^N$. Here we have identified $\RR^N\simeq (\RR^N)^*$. We will denote the components of these fields
$X^i(u)\in \Omega^0(D)$ and $\eta_j(u)\in \Omega^1(D)$.

The action is:
\[
S[X,\eta]=\int_D(\eta_i\wedge dX^i+\frac{1}{2} \omega^{ij}(X)\eta_i\wedge\eta_j)
\]
Here $\omega^{ij}(X)$ is the Poisson tensor corresponding to $\omega_{ij}$, i.e.
the inverse matrix to $\omega_{ij}(X)$. We will follow the agreement of summing up over
repeated contravariant and covariant indices.

As we will see below, the following boundary conditions for the field $(X,\eta)$
correspond to the amplitudes (\ref{PI-TQM}) in topological quantum mechanics:
\begin{itemize}
\item For $s\in I$, $(X,\eta)\in M\subset T^*M$ where $M$ is regarded as
a zero-section Lagrangian submanifold of $T^*M$, i.e. no condition on $X$ and $\eta|_I=0$.

\item For $s\in I_1$, $X(s)\in \cL_1$ and $\eta(s)\in N^*_{X(s)}\cL_1\otimes T^*_sD$ where $N^*\cL_1$ is the conormal bundle to $\cL_1$. Its fiber over $X(s)$ is the Lagrangian subspace in $T^*_{X(s)}M$, defined as $\{\eta(s)\in T^*_{X(s)}M|<a,\eta(s)>|_{I_1}=0\}$ for all $a\in T_{X(s)}\cL_1$. By restriction of
a form to the boundary we always mean the pull-back. Here $<-,->$ is the pairing of tangent and cotangent vectors,
\item Similarly, for $s\in I_2$, $X(s)\in \cL_2$ and $\eta(s)\in N^*_{X(s)}\cL_2\otimes T^*_sD$.
\item For $s\in I_{12}$, $X(s)\in \Lambda$ and $\eta(s)\in N^*_{X(s)}\Lambda\otimes T^*_sD$.
\end{itemize}

\subsubsection{The relation to topological quantum mechanics} The topological sigma model described above is related to the second order topological field theory introduced earlier as follows.
Consider the critical point of $S[X,\eta]$ in $\eta$. The functional is quadratic in $\eta$, therefore such a
critical point is unique and
it is given by
\begin{equation}\label{sel}
\overline{\eta}_i=-\omega_{ij}dX^j
\end{equation}
Let $V$ be an isotropic submanifold $M$. It is clear that if the boundary condition for $X$ is chosen in such a way  that $X(s)\in V$ when $s\in I\subset \pa D$
then $\overline{\eta}(s)\in N^*V$ when $s\in I$. Thus, this critical point agrees with boundary conditions above. Computing the action of the Poisson sigma model at the critical point for $\eta$ we recover the second order action for the $X$-field:
\[
S[X,\overline{\eta}]=\frac{1}{2} \int_D \omega_{ij}(X)dX^i\wedge dX^j
\]

Now let us describe the critical points of the action of Poisson sigma model in both $X$ and $\eta$. The Euler-Lagrange equations are
\[
dX^i+\omega^{ij}(X)\eta_j=0, \   \  d\eta_i+\frac{1}{2}\pa_i\omega^{kl}\eta_k\wedge\eta_l=0
\]
It is easy to see that the first equation implies the second one. Note that here we used the
invertibility of the Poisson tensor, i.e. the fact that the Poisson sigma model has a symplectic manifold
as the target space.

If Euler-Lagrange equations hold, the variation of the action is given only by boundary terms:
\[
\delta S[X,\eta]_{boundary}=\int_{\pa D}\eta_i\delta X^i
\]
This integral vanishes because of the boundary conditions. Indeed, $\delta X|_{I_a}\in T_X\cL_{a}$,
$\eta|_{I_a}\in N^*_X\cL_a$, which implies $\eta_i\delta X|_{I_a}=0$, similarly for $I_{12}$ with $\cL_a$ being replaced by $\Lambda$, and on $I$ the boundary term vanishes because $\eta=0$.

Thus, the critical points are given by (\ref{sel}) with an arbitrary smooth function $X:D\to M$ satisfying the boundary conditions described above.
Boundary conditions on $\eta$, together with the Euler-Lagrange equations, imply
\[
X(s)=c, \ \ s\in I, \ \ c\in \cL_1\cap \cL_2
\]
Thus, a critical point of $S[X,\eta]$ is a pair $(\overline{X},\overline{\eta})$ where $\overline{\eta}$ is given by (\ref{sel}) and $\overline{X}$ is any smooth function on $D$ satisfying the boundary conditions and $X|_I=c, c\in \cL_1\cap \cL_2$.

\subsubsection{Gauge invariance} The Poisson sigma model is gauge invariant with respect to the following
infinitesimal gauge transformations, i.e. vector fields on the space of fields
\[
\delta_\beta X^i=\omega^{ij}\beta_j, \ \ \delta_\beta \eta_i=-d\beta_i-\pa_i \omega^{jk}\eta_j\beta_k
\]
Here $\beta\in \Omega^0(D)\otimes \RR^N$ (in general $\beta(u)\in T^*_{X(u)}M$). The Lie subalgebra with
$\beta|_I=0$, $\beta(s)\in N^*_{X(s)}\cL_1$ when $s\in I_1$, $\beta(s)\in N^*_{X(s)}\cL_2$ when $s\in I_2$ and $\beta(s)\in N^*_{X(s)}\Lambda$ when $s\in I_{12}$ preserves the
boundary conditions corresponding to the topological quantum mechanics.

The gauge symmetry of the Poisson sigma model is directly related to the
reparametrization invariance of topological quantum mechanics.

\subsection{Amplitudes} Thus, up to constants we expect the following formula for scalar products (amplitudes) in terms of the Poisson sigma model:
\[
(\psi^{(2)}_{b_2}, \psi^{(1)}_{b_1})=\int_{(X,\eta):D\to T^*M} e^{\frac{i}{h}S(X,\eta)} DXD\eta
\]
Here we assume the boundary conditions described in section \ref{PSM-bc} and the path integral should be understood semiclassically, as a formal power series of
Feynman diagrams. Because the action is gauge invariant and therefore highly degenerate,
to make sense of such a path integral we should do the gauge fixing. We will do it later
using Lorenz gauge fixing and an extra gauge fixing on the boundary.

As it was discussed before, changing the auxiliary Lagrangian submanifold $\Lambda$ will only produce
a constant $U(1)$-factor.

\subsection{Cyclic amplitudes}
We can do the same to the path integral describing cyclic amplitudes as we did for amplitudes.
This gives the following representation of cyclic amplitudes in
terms of the Poisson sigma model:
\begin{equation}\label{cycl-psm}
(\psi^{(1)}_{b_1}, \psi^{(k)}_{b_{k}}) \dots (\psi^{(3)}_{b_3}, \psi^{(2)}_{b_2}) (\psi^{(2)}_{b_2}, \psi^{(1)}_{b_1})=
\int_{(X,\eta):D\to T^*M} e^{\frac{i}{h}S(X,\eta)} DXD\eta
\end{equation}
Here we impose boundary conditions
\begin{itemize}
\item $\eta=0$ and $X$ is not fixed on $J_a$,
\item $X(s)\in \cL_a$ and $\eta(s)\in N^*_{X(s)}\cL_a\otimes T^*_sI_a$, when $s\in I_a$.
\end{itemize}

The natural question about the formula (\ref{cycl-psm}) is why does the path integral factorize into the
product of path integrals describing the amplitudes? Roughly speaking, this happens because the field theory is
topological. This implies that we can deform the disc on Fig.\ref{cyclic} into a semi-infinite strip with the side
corresponding to the auxiliary Lagrangian $\Lambda$ (which we have to choose)  at infinity. In terms of the gauge fixing this
means choosing a very special metric on $D$. In this limit the critical points are simply $X(u)=c\in \cL_1\cap \cL_2$
and $\eta=0$. Feynman diagrams vanish if any of their edges extend to infinity. Thus, all Feynman diagrams
are supported at the end of the semi-infinite strip near the boundary segment $I$.

\begin{figure}[htb]
\includegraphics[height=6cm,width=6cm]{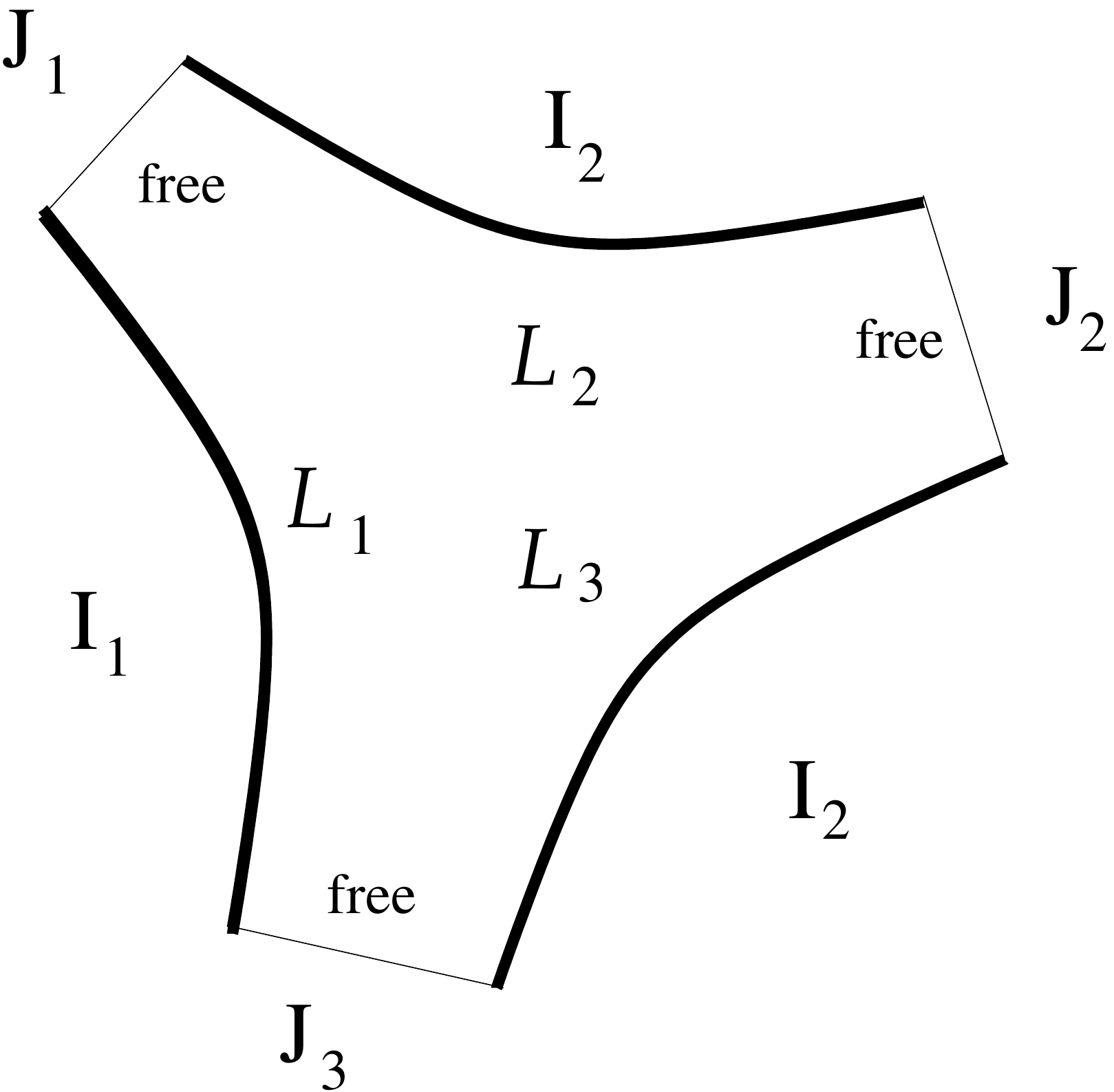}
\caption{Deformation of the disc into a star with infinitely long tentacles for $k=3$.}
\label{Def}
\end{figure}

For the same reasons, a disc with alternating boundary conditions (corresponding to cyclic
amplitudes) can be deformed into a ``star" with semi-infnite ``tentacles", as it is shown on Fig. \ref{Def}. In this
limit critical points can be chosen as constant maps $X(u)=c\in \cL_a\cap \cL_{a+1}$ near the end of the star's tentacles. Also, in this limit only the Feynman diagrams localized near the ends of the star will contribute and these contributions are exactly the amplitudes of the topological quantum mechanics discussed above. Thus, in this limit the path integral (\ref{cycl-psm}) factorizes as expected. But since the field theory is topological, this factorization is exact.

This is the heuristics of the factorization. The factorization in terms of Feynman diagrams involves
direct manipulations with the diagrams (as in case of the associativity of the star product in \cite{K})
and will be done in a separate publication.

\section{Classical BV extended Poisson sigma model and gauge fixing}

\subsection{Classical BV-BFV Poisson sigma model}Recall that the BV extended Poisson sigma model is an AKSZ model where fields on $D$ which can be naturally organized
in supermultiplets \cite{CF}:
\[
\widetilde{X}^i=X^i+\eta^{\dag,i}+\beta^{\dag,i}, \ \ \widetilde{\eta}_i=\beta_i+\eta_i+X^\dag_i
\]
The space of fields $\cF_D$ on the bulk $D$  is $\ZZ$-graded by the ghost number: $gh(X)=gh(\eta)=0$, $gh(\eta^\dag)=gh(X^\dag)=-1$,
$gh(\beta)=1$, $gh(\beta^\dag)=-2$. The fields are differential forms $X,\beta\in \Omega^0(D, \RR^N)$, $\eta, \eta^\dag\in \Omega^1(D,\RR^N)$
and $\beta^\dag, X^\dag\in \Omega^2(D,\RR^N)$. The space of fields $\cF_{\pa D}$ on the boundary $\pa D$ consists of pull backs of $X, \eta, \eta^\dag, \beta$.

The space of fields on $D$ can be naturally regarded as a shifted cotangent bundle to the space of fields
$X$, $\eta$ and $\beta$.

Both $\cF_D$ and $\cF_{\pa D}$ have natural symplectic structures:
\[
\omega_D=\int_D \delta X^\dag_i\wedge \delta X^i+ \int_D \delta\eta^{\dag,i}\wedge\delta\eta_i+\int_D\delta\beta^{\dag,i}\wedge\delta\beta_i
\]
\[
\omega_{\pa D}=\int_{\pa D} \delta \eta_i\wedge \delta X^i+ \int_D \delta\beta_i\wedge\delta\eta^{\dag, i}
\]

The action functional of the BV extended
Poisson sigma model is
\begin{equation}\label{BVP}
S_D=\int_D(\widetilde{\eta}_i\wedge d\widetilde{X}^i+\frac{1}{2} \omega^{ij}(\widetilde{X})\tilde{\eta}_i\wedge\tilde{\eta}_j)
\end{equation}
The boundary action
\[
S_{\pa D}=\int_{\pa D}(\tilde{\eta}_i\wedge d\tilde{X}^i+\frac{1}{2} \omega^{ij}(\tilde{X})\tilde{\eta}_i\wedge\tilde{\eta}_j)=
\int_{\pa D} \beta_i dX^i+ \omega^{ij}(X)\beta_i\eta_j+\frac{1}{2}\pa_k\omega^{ij}(X)\eta^{\dag,k}\beta_i\beta_j
\]
together with cohomological vector fields
\[
Q_D=\int_D (d\widetilde{X}^i+\omega^{ij}(\widetilde{X})\wedge \widetilde{\eta}_j)\wedge \frac{\delta }{\delta \widetilde{X}^i}+\int_D(d\widetilde{\eta}_i+ \frac{1}{2}\pa_i\omega^{kl}(\widetilde{X})\wedge
\widetilde{\eta}_k\wedge \widetilde{\eta}_l)\wedge \frac{\delta }{\delta \widetilde{\eta}_i}
\]
and
\[
Q_{\pa D}F=\{S_{\pa D}, F\}_{\pa D}
\]
defines the BV-BFV structure of the models  \cite{CMR}.

Now let us discuss boundary conditions for the BV extended model.

\subsection{Boundary conditions corresponding to the topological quantum mechanics}
Boundary conditions for BV extended field theories are $Q_\partial$-invariant Lagrangian submanifolds in the space of
boundary BFV fields \cite{CMR}.

Boundary conditions corresponding to amplitudes in topological quantum mechanics are:

\begin{itemize}

\item On $I$, no conditions on $\tilde{X}$ and $\tilde{\eta}=0$, i.e. fields are in the zero section Lagrangian.

\item When $s\in I_1$, $X(s)\in \cL_1$, $\eta^\dag(s)|_{I_1}  \in T_{X(s)}[-1]\cL_1\otimes T^*_sI_1$,     $\eta(s)|_{I_1} \in N^*_{X(s)}\cL_1\otimes T^*_sI_1$  , and $\beta(s)|_{I_1}  \in N^*_{X(s)}[1]\cL_1\otimes T^*_sI_1$.

\item When $s\in I_2$, $X(s)\in \cL_2$, $\eta^\dag(s)|_{I_2}  \in T_{X(s)}[-1]\cL_2\otimes T^*_sI_2$,     $\eta(s)|_{I_2} \in N^*_{X(s)}\cL_2\otimes T^*_sI_2$  , and $\beta(s)|_{I_2}  \in N^*_{X(s)}[1]\cL_2\otimes T^*_sI_2$.

\item And similarly, when $s\in I_{12}$, $X(s)\in \Lambda$, $\eta^\dag(s)|_{I_{12}}  \in T_{X(s)}[-1]\Lambda\otimes T^*_sI_{12}$,     $\eta(s)|_{I_{12}} \in N^*_{X(s)}\Lambda\otimes T^*_sI_{12}$  , and $\beta(s)|_{I_{12}}  \in N^*_{X(s)}[1]\Lambda\otimes T^*_sI_{12}$.

\end{itemize}
Here, as before $\cL_a=\cL_{b_a}^{(a)}$.
Note that boundary conditions for $\beta$ and $\eta^\dag$ do not follow from boundary conditions for
$X$ and $\eta$. Our choice of boundary conditions is such that the boundary Lagrangian submanifold is tangent
to the vector field $Q_{\pa D}$. In other words gauge transformations preserve this boundary condition.

Boundary conditions corresponding to cyclic amplitudes on intervals $J_a$ are the same as above on $I$,
and on intervals $I_a$, are the same as above for $I_1$,
$I_2$ or $I_{12}$ for corresponding Lagrangian submanifolds.
For more details see  \cite{CF1}\cite{CF2}.

\subsection{Gauge fixing}
Recall that in the BV framework gauge fixing means a choice of a Lagrangian submanifold
in the odd-symplectic space of bulk fields. In case when the gauge symmetry is the
result of a gauge group action, gauge fixing is a choice of a section of the
projection from the space of fields to gauge classes of fields. This is equivalent to choosing a Lagrangian
submanifold in a BV extended field theory. In the case of Poisson sigma model, where the gauge
symmetry is not given by a group action, choosing such a Lagrangian submanifold
is the analog of choosing a gauge fixing section.

We choose the Lorenz gauge fixing which means we fix the
following Lagrangian submanifold in $\cF_D$:
\[
L_{gf}=\{X^{\dag}_{i}=\beta^{\dag,i}=0, \ \ \eta^{\dag,i}=*d\gamma^i, \ \ d^*\eta_i=0\}
\]
In particular, this means  $d^*\widetilde{\eta}_i=0$ and $d^*\widetilde{X}^i=0$. Because $D$ is simply-connected, the condition $d*\eta_i=0$ implies that $\eta_i=*df_i$ for some function $f$.

Let us check that this subspace with the boundary conditions corresponding to the topological quantum mechanics is indeed Lagrangian. Consider the restriction of the symplectic form
\[
\omega_D=\int_D \delta X^{\dag}_i\wedge \delta X^i+ \int_D \delta\eta^{\dag,i}\wedge\delta\eta_i+\int_D\delta\beta^{\dag,i}\wedge\delta\beta_i
\]
to $L_{gf}$. The first term is zero because $X^\dag=0$ on $L_{gf}$, the third term is zero because on this submanifold $\beta^\dag=0$. The second term restricted to $L_{gf}$ can be written as :
\[
\int_D\delta \eta^{\dag,i}\wedge \delta\eta_i=\int_D*d\delta\gamma^i\wedge\delta\eta_i=-\int_D\delta\gamma^i d*\delta\eta_i+\int_{\pa D}\delta\gamma^i\ast\delta\eta_i
\]
The first integral is zero on $L_{gf}$. In order for the boundary term in this integral  to vanish, one needs to put an extra boundary condition
$\gamma(s) \in  T_{X(s)} \cL_a$   for $s \in I_a$
and $\gamma=0$ on $I$ (where $\eta=0$).
In particular, this condition implies that the normal component of $\eta^+$ is in $T_{X(s)} \cL_a$ for $s \in I_a$ and vanishes for $s \in I$\footnote{In order for the space of fields, subject to the boundary conditions corresponding to topological quantum mechanics to be closed under $d$ and $d^*$, one needs to refine (or ``strictify") the boundary conditions.

On $I$, one should impose the ultra-Dirichlet boundary condition for $\tilde{\eta}$ and ultra-Neumann for $\tilde{X}$, cf. \cite{CMR2}, Appendix A.1 (see also Appendix A.3.2 in the same reference for the logic of strictification of the boundary conditions). To refine boundary conditions for $s\in I_a$ one should split $T^*M$ into the direct sum of $N^*\cL_a$ and 
a complementary vector space. This can be achieved  by choosing an almost complex structure $J$ on $TM$. The refined boundary conditions can be chosen then as
even normal derivatives of  $\tilde{X}|_{I_a}$ and odd normal derivatives of  $(* \tilde{X} ) |_{I_a}$ are in $T \cL_a$;
odd normal derivatives of $\tilde{X}  |_{I_a}$ and even normal derivatives of $(*\tilde{X})  |_{I_a}$ are in $J (T \cL_a)$;
even normal derivatives of  $\tilde{\eta} |_{I_a}$  and odd normal derivatives of $( * \tilde{\eta})  |_{I_a}$ are in $N^*  \cL_a$;
odd normal derivatives of $\tilde{\eta} |_{I_a}$ and even normal derivatives of $(*\tilde{\eta}) |_{I_a}$ are in $J^* (N^*  \cL_a)$.

To be more precise, we need these boundary conditions for the fluctuations around classical solutions, but we will defer this discussion to
the extended version of this note.}.

The gauge fixing submanifold $L_{gf}$ is a Lagrangian submanifold of conormal type if we will impose the ``strictification" of boundary
conditions discussed above. One can define an equivalent gauge fixing 
using the ``gauge fixing fermion". For this one needs to introduce a quadruplet of auxiliary fields $\gamma$, $\overline{\nu}$ (Lagrange multiplier), $\gamma^\dag$ and $\nu^\dag$. Such extra fields are typical for the BV extended version of Faddeev-Popov gauge fixing with Lagrange multipliers. For details, see \cite{CF}.

\subsection{Solutions to the Euler-Lagrange equations} The gauge fixed action is the restriction
of the action to $L_{gf}$ which was computed in \cite{CF}:
\[
S_{gf}[\tilde{X},\tilde{\eta}]=\int_D(\eta_i\wedge dX^i+\frac{1}{2} \omega^{ij}(X)\eta_i\wedge \eta_j-
\eta^{\dag,i}\wedge (d\beta_i+\pa_i\omega^{kl}(X)\eta_k\beta_l)-\frac{1}{4}\eta^{\dag,i}\wedge\eta^{\dag,j}\pa_i\pa_j\omega^{kl}\beta_k\beta_l)
\]
In this action $\eta_i=*df_i$ and $\eta^{\dag, i}=*d\gamma^i$.
For the variation we have:
\begin{equation}\label{varS}
\delta S_{gf}[\tilde{X},\tilde{\eta}]=\mbox{ bulk terms} +\int_D (\eta_i\delta X^i+\eta^{\dag, i}\delta\beta)
\end{equation}
As usual, it consists of two terms: the bulk term and
the boundary term. The bulk terms vanishes if and only if the Euler-Lagrange equations hold:
\[
d^*(dX^i+\omega^{ij}(X)\eta_j-\eta^{\dag, l}\wedge \pa_l\omega^{ij}(X)\beta_j)=0,
\]
\[
 d\eta_i+\frac{1}{2}\pa_i\omega^{kl}\eta_k\wedge\eta_l-\eta^{\dag, l}\wedge \pa_l\pa_i\omega^{jk}(X)
\eta_j\beta_k-\frac{1}{4}\eta^{\dag, k}\wedge\eta^{\dag, l}\pa_k\pa_l\pa_i \omega^{st}\beta_s\beta_t=0
\]
\[
d^*(d\beta_i+\pa_i\omega^{jk}(X)\eta_j\beta_k+\frac{1}{2}\eta^{\dag,j}\pa_i\pa_j\omega^{km}(X)\beta_k\beta_m)=0,
\]
\[ d\eta^{\dag, i}+\eta^{\dag, j}\wedge \pa_j\omega^{ki}(X)\eta_k
+\frac{1}{2} \eta^{\dag, l}\wedge\eta^{\dag, j}\pa_l\pa_j\omega^{ki}(X)\beta_k=0
\]
The boundary terms in (\ref{varS}) vanish because of the boundary conditions described earlier. These equations with boundary conditions described earlier still have infinitely many solutions
because the gauge symmetry on the boundary is not yet fixed, see the discussion below.

\subsection{The case of constant symplectic structure on $\RR^{2n}$}
From now on we will focus on the simplest but already non-trivial case when $M=\RR^{2n}$ with constant symplectic structure. In this case the Euler-Lagrange equations become:
\[
d^*(dX^i+\omega^{ij}\eta_j)=0, \qquad d\eta_j=0, \qquad d^*d\beta_i=0, \qquad d\eta^{\dag, i}=0
\]
We also should take into account the gauge fixing conditions $\eta^{\dag, i}=*d\gamma^i,  d^*\eta=0$
and the boundary conditions. 
This gives
\begin{enumerate}
\item
\[
d^*dX^i=0, \qquad X|_{I_\alpha}\in \cL_\alpha, \qquad X_{I_{12}}\in \Lambda, \qquad dX^i|_I=0
\]
Here and below $\alpha=1,2$.

\item For $\eta$ we have
\[
d\eta=0
\]
with boundary conditions $\eta|_I=0$, $\eta(s)|_{I_\alpha}\in N^*_{X(s)}\cL_\alpha\otimes T^*_sI_\alpha$ when $s\in I_\alpha$ and $\eta(s)|_{I_{12}}\in N^*_{X(s)}\Lambda\otimes T^*_sI_{12}$ when $s\in I_{12}$.

\item  For $\beta_i$ we have the linear boundary value problem
\[
d^*d\beta_i=0, \qquad \beta_i|_I=0,
\]
with $\beta(s)|_{I_\alpha}\in N^*_{X(s)}\cL_\alpha\otimes T^*_sI_\alpha$ when $s\in I_\alpha$, $\beta(s)|_{I_{12}}\in N^*_{X(s)}\Lambda\otimes T^*_sI_{12}$ when $s\in I_{12}$, and $\beta|_I=0$.

\item For fields $\gamma^i$ we obtain linear problem:
\[
d^*d\gamma^i=0
\]
with boundary conditions $*d\gamma(s)|_{I_\alpha}\in T_{X(s)}\cL_\alpha\otimes T^*_sI_\alpha$ when $s\in I_\alpha$,
$*d\gamma(s)|_{I_{12}}\in T_{X(s)}\Lambda\otimes T^*_sI_{12}$ when $s\in I_{12}$, and free boundary conditions
on $I$.

\end{enumerate}

Note that Euler-Lagrange equations are linear, but boundary conditions are not.
It is clear that these boundary conditions are incomplete: we still have infinitely many solutions to Euler-Lagrange equations
with these boundary conditions. This is due to remaining gauge symmetry on the boundary.
To observe the remaining gauge symmetry let us focus on fields with $gh=0$, i.e. on $X$ and $\eta$. When the symplectic structure
is constant, initesimal gauge
transformations are
\[
\delta X^i=\omega^{ij}\beta_j, \ \ \delta\eta_i=-d\beta_i
\]
Such a vector field is parallel (tangent) to the Lorenz gauge fixing section $d*\eta=0$ only if $d^*d\beta_i=0$,
i.e. when $\beta$ is harmonic. Boundary values of $\beta$ are constrained by conditions on $X$. Since $X(s)\in \cL_\alpha$
when $s\in I_\alpha$ we have $\delta X(s)\in T_{X(s)}\cL_\alpha$ and therefore $\beta \in N^*_{X(s)}\cL_\alpha$. This agrees
with boundary conditions on $\eta$. However, there are infinitely many such harmonic functions. They form the remaining boundary gauge symmetry.
We call it boundary gauge symmetry because a harmonic function is determined by its boundary value.

It is easy to see that the refined boundary conditions discussed in the previous section fix this boundary gauge symmetry.

\section{Quantization of the BV extended Poisson sigma model}

Here we will focus on the Poisson sigma model with boundary conditions corresponding to
amplitudes and cyclic amplitudes in the topological quantum mechanics.

According to the general philosophy of BV quantization, we should choose a gauge fixing Lagrangian submanifold
in the space of bulk BV fields in such a way that the BV extended action has isolated critical points on this Lagrangian.
Then the perturbative partition function is defined by the sum of corresponding Feynman diagrams. Thus,
our main object, should be

\[
\int_{L_{gf}} \exp(\frac{i}{h}S_{gf}[\widetilde{X}, \widetilde{\eta}]) D\widetilde{X}D\widetilde{\eta}
\]
with boundary conditions described above.

However, in our case Lorenz gauge fixing does not provide isolated critical points because of the remaining gauge
transformations which are parallel to the Lagrangians $\cL_a$ (unless we consider refined boundary conditions). 
In addition to this, we have nonlinear boundary conditions.
One way to tackle both problems is to treat the boundary conditions as boundary constraints using Lagrange multipliers.

The main idea in the rest of this section is to introduce Lagrange multipliers $\lambda$ for the boundary constraints on $X$, keep $X$ free on the boundary and impose boundary conditions $\eta=0$\footnote{To be more precise, we should do this only to
fluctuations, but we will discuss this, as many other missing details in this note in its extended version.}.
The usual mechanism with Lagrange multipliers will
constrain field $X$ to $\cL_a$ on $I_a$. Lagrange multipliers $\lambda$ parameterize the ``true" boundary values of $\eta$.
Thus, Lagrange multipliers will effectively change boundary condition $\eta=0$ and free $X$ on each $I_a$ to the correct boundary
condition corresponding to the Lagrangian submanifold $\cL_a$. We conjecture that this gives correct Feynman diagram expansion in the path integral. Here we consider the case of the amplitude in the asymptotic metric, so $a=1,2$.

Assume that Lagrangian fibers  $\cL^{(a)}_{b_a}$ are the level surfaces of the Poisson commuting  functions
$H^{(a)}_\alpha\in C(M)$, where $\alpha=1,\dots, dim(M)/2$. This is a reasonable assumption  in the context of integrable systems where these functions are commuting Hamiltonians.

In the BV framework we want to introduce an extra term $\widetilde{S}_\pa$ in the action with Lagrange
multipliers and their BV counterparts in such a way that the total action satisfies the classical master equation.
In order to do this, introduce boundary fields $\lambda^a_\alpha \in \Omega^1(I_a),  \lambda^{\alpha,\dag}_a\in \Omega^0(I_a)[-1]$, $c_\alpha^a \in \Omega^0(I_a)[1],  c^{\alpha,\dag}_a \in \Omega^1(I_a)[-2]$, $\overline{c}_a^\alpha \in \Omega^1(I_a)[-1], \overline{c}^{a,\dag}_\alpha \in \Omega^0(I_a)$, $\mu^\alpha_a \in \Omega^1(I_a),  \mu^{a,\dag}_\alpha\in \Omega^0(I_a)[-1]$. Denote this space of fields $\widetilde{F}_{\pa D}$. Naturally
\[
\widetilde{F}_{\pa D}=\oplus_a \widetilde{F}_{I_a}
\]
Each space $\widetilde{F}_{I_a}$ has a natural symplectic structure of degree $-1$:
\[
\widetilde{\omega}_{I_a}=\int_{I_a}\sum_{\alpha}( \delta\lambda^{\alpha,\dag}_a\wedge \delta\lambda_\alpha^a+
\delta c^{\alpha,\dag}_a\wedge \delta c_\alpha^a+\delta\overline{c}^{a,\dag}_\alpha\wedge \delta\overline{c}_a^\alpha+\delta\mu^{a,\dag}_\alpha\wedge \delta\mu_a^\alpha)
\]

Note that the symplectic space $(\widetilde{F}_{\pa D}, \widetilde{\omega}_{\pa D})$ is not the space of BFV fields,
but the space of BV extended Lagrange multipliers.

Denote by $F^{(0)}_D$ the space of bulk fields with boundary conditions $\eta|_{\pa D}=0$ and $X|_{\pa D}$ is free
and define the total space of fields as  $\cF_D=F^{(0)}_D\oplus \widetilde{F}_{\pa D}$. This space has a natural
symplectic form of degree $-1$ which is the direct sum of symplectic forms on $F^{(0)}_D$ and $\widetilde{F}_{\pa D}$.

Define the boundary correction to the action as
\[
\widetilde{S}_\pa=\sum_a\sum_\alpha \int_{I_a} (\lambda_\alpha^a (H_a^{(\alpha)}-b^{\alpha}_a)+\lambda^{\alpha,\dag}_adc_\alpha^a+c_\alpha^a\pa_iH^{(\alpha)}_a\eta^{i,\dag}+
\mu_a^\alpha\overline{c}^{a,\dag}_\alpha)
\]
Again, this is not the BFV action, this is a BV extended term with Lagrange multiplier enforcing the boundary conditions on $X$.

It is easy to check that the modified action $S+\widetilde{S}_\pa$ satisfies the classical master equation.
\[
\{ S+\widetilde{S}_\pa, S+\widetilde{S}_\pa \}_D=0
\]
where the Poisson bracket is defined by the symplectic structure on the total space of BV fields $\cF_D$.

We will choose Lorenz gauge fixing in $F^{(0)}_D$, as in \cite{CF}. The additional gauge fixing for BV Lagrange multipliers can be chosen as a version of the Lorenz gauge fixing.
Corresponding Lagrange submanifold in $\widetilde{\cF}_\pa$ is the graph of the function
\[
\psi=\sum_a \int_{I_a}\sum_\alpha \overline{c}^\alpha_a d^*\lambda_\alpha^a
\]
The Lagrangian submanifold is defined as $a^\dag=\frac{\delta \psi}{\delta a}$ where $a$ are boundary fields
which translates to:
\[
\lambda^{\alpha,\dag}_a=-d^*\overline{c}^\alpha_a, \ \
c^{\alpha,\dag}_a=0
\]
\[
\overline{c}^{a,\dag}_\alpha=d^*\lambda^{a}_\alpha, \ \
\mu^{\alpha,\dag}_a=0
\]

For the restriction of the boundary action to this Lagrangian submanifold we have
\[
\widetilde{S}_{\pa, gf}=\sum_a\sum_\alpha \int_{I_a} (\lambda_\alpha^a (H_a^{(\alpha)}-b^{\alpha}_a)-d^*\overline{c}^\alpha_adc_\alpha^a+c_\alpha^a\pa_iH^{(\alpha)}_a\eta^{i,\dag}+
\mu_a^\alpha d^*\lambda_{\alpha}^a )
\]
Note that if we ignore all boundary fields except $\lambda$, Euler-Lagrage equations for $S+\widetilde{S}_{\pa D}$ induce
Euler-Lagrange equations discussed before in the bulk. On the boundary, on $I_a\in \pa D$  we will have $H_a^{(\alpha)}(X)=b_a^\alpha$ which means
$X\in \cL_a$ and we will also have $\eta_i=\sum_\alpha \lambda_\alpha \pa_i H_a^{(\alpha)}$ which means $\eta(s)\in N^*_{X(s)}\cL_a$.

\subsection{Gluing}

Two Lagrangian fibrations $\pi_a: M\to B_a$  on $M$ define two real polarizations $P_1, P_2$. Denote by $U_{P_2,P_1}: H_{P_1}^{1/2}\to H_{P_2}^{1/2}$ the integral operator with the integral kernel $(\psi^{(2)}_{b_2},\psi_{b_1}^{(1)})$. Here by an
integral operator we mean a formal integral operator where the integration is given by the formal power series in Feynman diagrams. This is a semiclassical version of the Blattner-Kostant-Sterberg kernel in geometric quantization.

Path integral arguments naturally lead to the composition rule
\[
U_{P_2,P}*U_{P,P_1}=U_{P_2,P_1}\exp(\frac{i\pi }{4}\mu(P_1,P,P_2))
\]
where $\mu(P_1,P,P_2)$ is the Maslov index. Here the composition is
the formal stationary phase integral over $B$ given by contributions from critical points.
We conjecture that this identity holds as an identity of formal powers series
with coefficients being Feynman diagrams for the Poisson sigma model with
appropriate boundary conditions. In terms of the path integral, it is a two-step
argument. First because our field theory is  topological, we can deform the gauge fixing
choice of the metric to a metric where the interval $I$ shrinks to a point. Then
integrating over $B$ allows all possible values of $X$ on the interval $I$
and thus, removes the constraint.

\begin{figure}[htb]
\includegraphics[height=4cm,width=8cm]{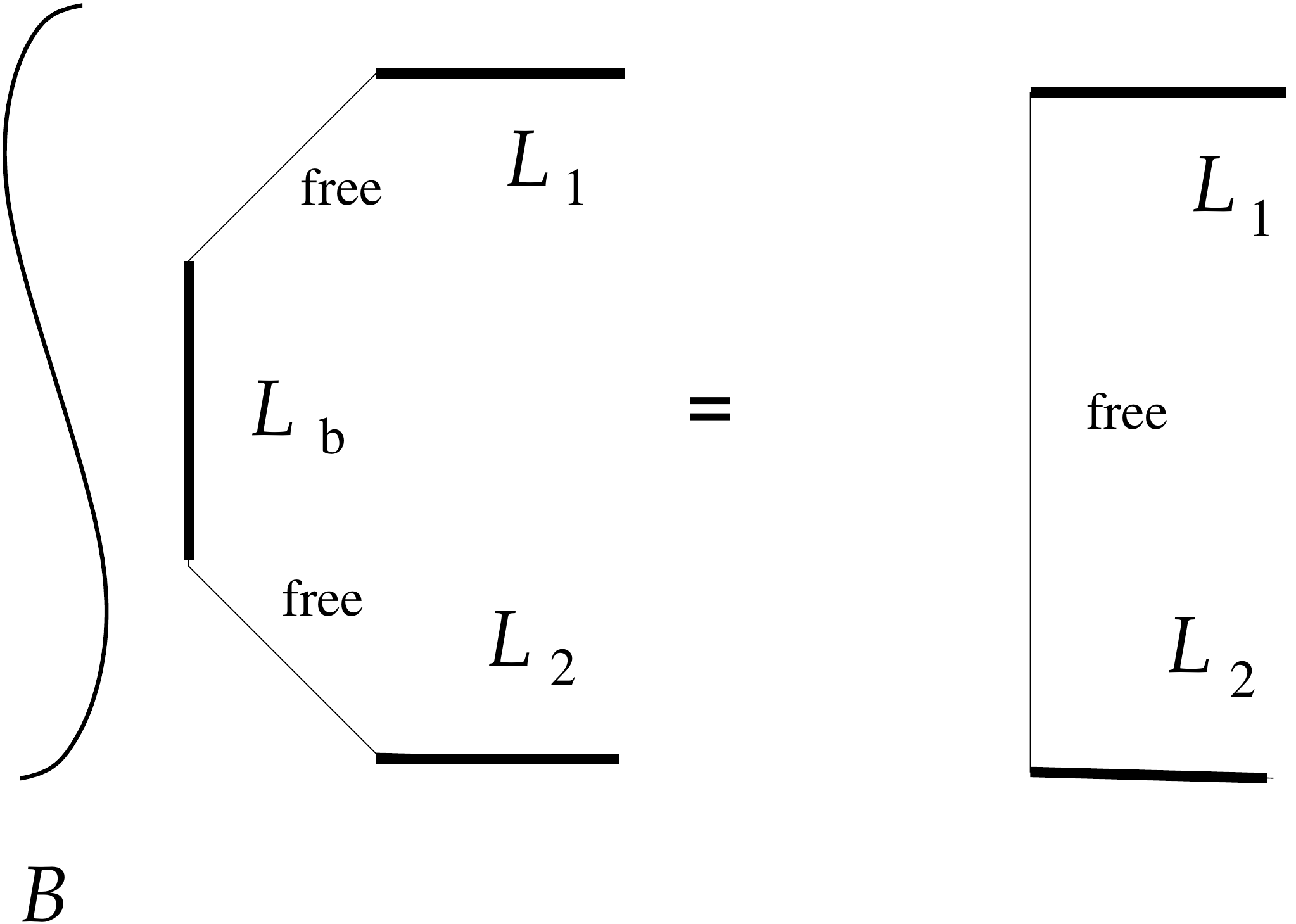}
\caption{Composing two BKS kernels.}
\label{GL}
\end{figure}

Pictorially this identity corresponds to gluing identity from Fig. \ref{GL}.

\section{Geometric representation of Kontsevich's star products for Poisson manifolds}

Here we will denote by $\cP$ a Poisson manifold, i.e. a smooth manifold with
a bivector field $p\in \Gamma(\wedge^2 T^*\cP)$ such that the bracket
\[
\{f,g\}=p(df\wedge dg)
\]
defines a Lie algebra structure on the space $C(\cP)$  of smooth functions on $\cP$.

Recall that a star product of two smooth functions on $\cP$ is an associative
product on $C(\cP)[[h]]$ defined as

\[
f*g=fg+\frac{ih}{2}\{f,g\}+\sum_{n\geq 2} m_n(f,g) (ih)^n
\]
where coefficients $m_n$ are bidifferential operators. The associativity of the star product
is equivalent to certain bilinear identities for coefficients $m_n$.
Kontsevich in \cite{K} gave an explicit construction of such
star product where the bidifferential operators $m_n$ are described explicitly  in terms of
integrals on configuration spaces.

\subsection{Kontsevich's star-product as Poisson sigma model}
In this section, as above, we assume that $D$ is an oriented disc.
We think of $D$ as a disc in $\RR^2$ and assume the counter clock wise
orientation.

\begin{figure}[htb]
\includegraphics[height=5cm,width=5cm]{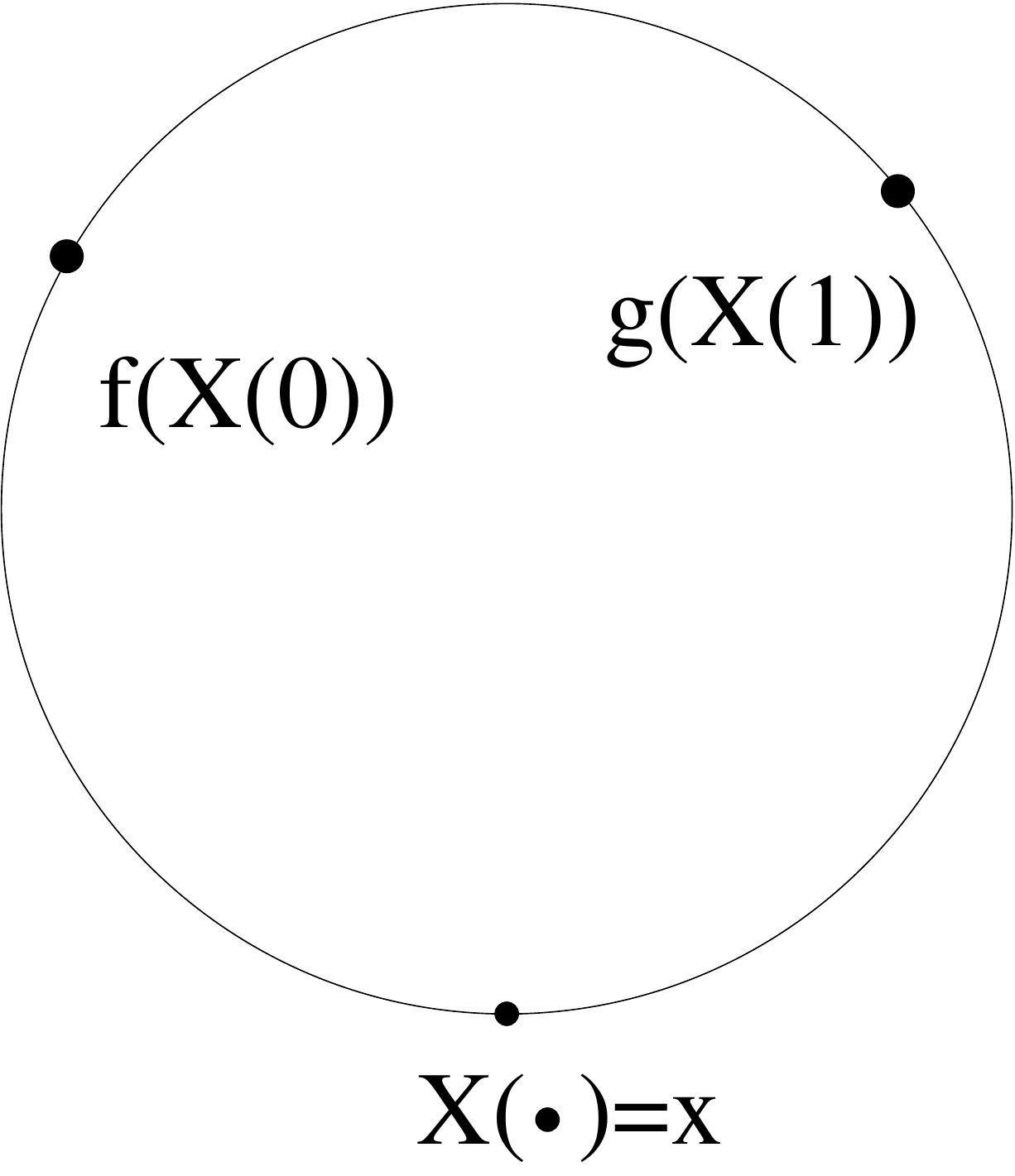}
\caption{The disc with three points $\bullet, 0, 1$  on the boundary and
insertions of functions $f$ and $g$ evaluated at $X(0)$ and $X(1)$ in the integral.}
\label{F-P1}
\end{figure}

The Poisson $\sigma$-model with boundary conditions $\eta=0$ and $X$  free \cite{CF}
describes Kontsevich's star product \cite{K} on the space of smooth functions on a
Poisson manifold $\cP$. Recall this construction. The star product of two smooth functions $f$ and $g$
can be regarded as the path integral
\[
f*g=\int_{X(\bullet)=x} \exp(\frac{i}{h}S[X,\eta]) f(X(0))g(X(1)) DXD\eta
\]
Here $\bullet, 0, 1$ are three points on the boundary of the disc (see Fig. \ref{F-P1}).
The integral is understood as a formal power series of Feynman integral contributions
from fluctuations around the critical point $X(u)=x, \eta(u)=0$. The description of
Feynman diagrams in the Lorenz gauge in BV framework was derived in \cite{CF}.

\begin{figure}[htb]
\includegraphics[height=4cm,width=8cm]{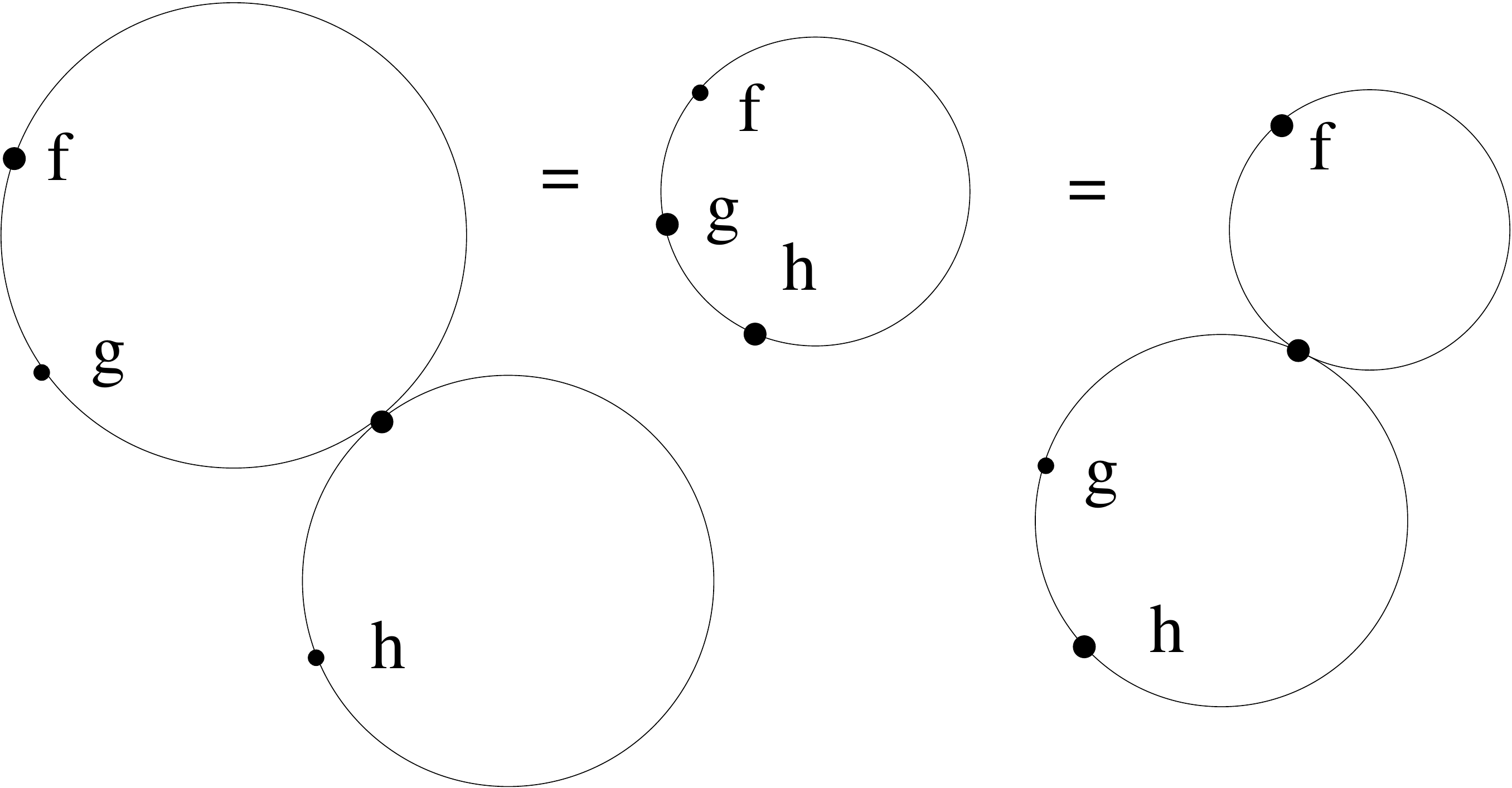}
\caption{The path integral interpretation of the associativity.}
\label{F-P2}
\end{figure}

The associativity of this star product can be proven directly as in \cite{K}.
It has a simple heuristic path integral interpretation. The associativity of the star product can
be interpreted  as the computation of the
path integral in two different ways as it is shown on Fig. \ref{F-P2} where all path integrals
should be understood perturbatively, i.e. as formal series in $h$ with coefficients given by Feynman diagrams.
Note that, of course, this path integral interpretation should be regarded only as a guideline for how to prove
the identity between the Feynman diagrams, but not as a proof itself. The proof remains
to be the direct proof given in \cite{K}.

Making formal deformations of $\omega$ in Kontsevich's start product one can obtain all non-equivalent $*$-products.
Changing coordinates on the target space by formal diffeomorphisms we obtain all equivalent star products. 

\subsection{Poisson sigma model with Lagrangian boundary conditions}
Let $S\subset \cP$ be a symplectic leaf. Assume we have three Lagrangian
fibrations on $S$, $\pi: S\to B$ and $\pi_{1,2}:S\to B_{1,2}$.

Thus, as in the first section, we can define the vector space of
$1/2$-densities $H_P^{1/2}$ and two integrable systems corresponding to
two other Lagrangian fibrations. We want to define the action of
the associative algebra $(C(\cP)[[h]], *)$ on the vector space $H_P^{1/2}$. That is, to each $f\in C(\cP)[[h]]$
we want to assign $\pi(f): H_P^{1/2} \to H_P^{1/2}$, in such a way that $\pi(f*g)=\pi(f)\pi(g)$.

\begin{figure}[htb]
\includegraphics[height=4cm,width=8cm]{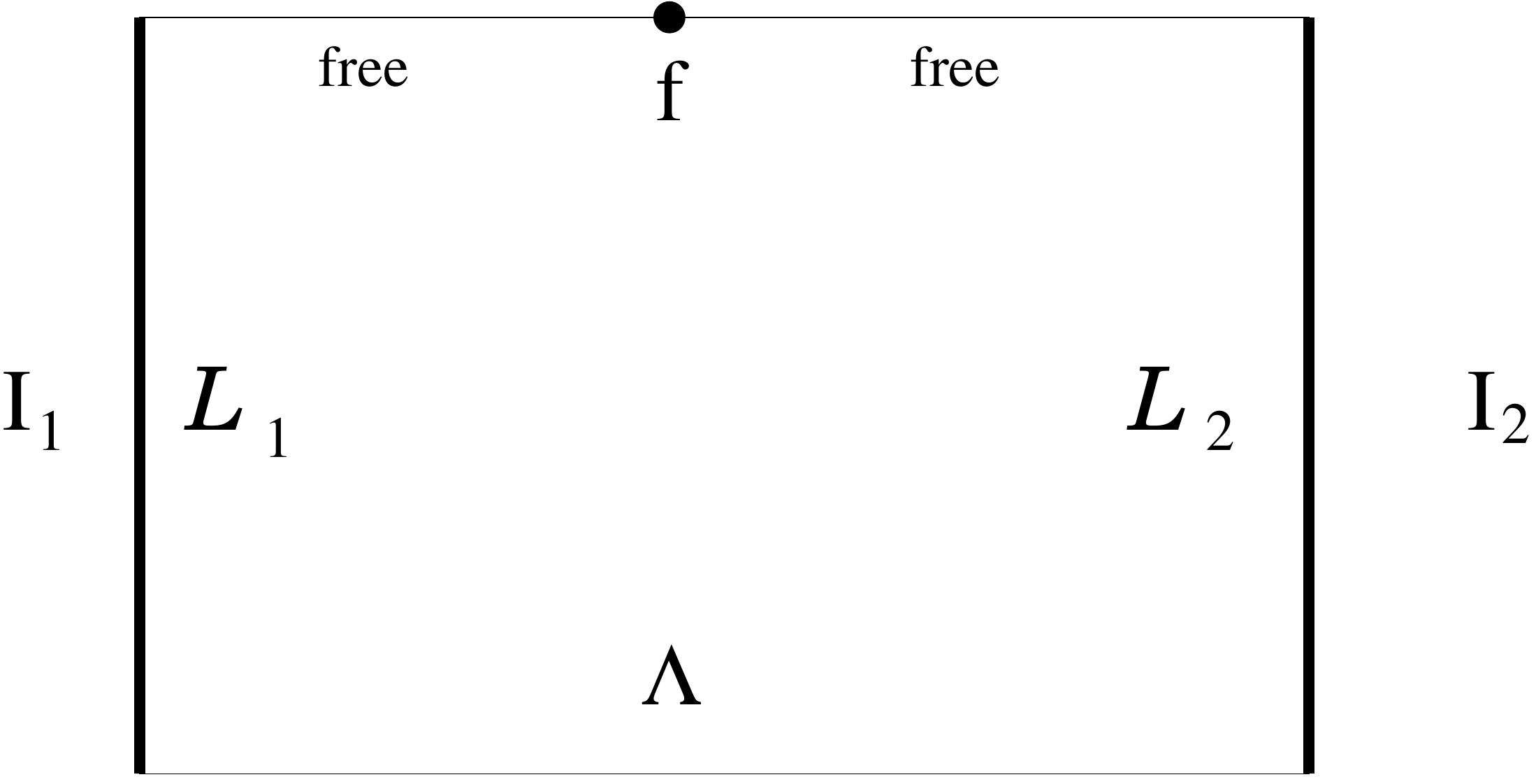}
\caption{The boundary conditions for the path integral representing matrix elements $(\psi_{b_2}^{(2)}, \pi(f)\psi^{(1)}_{b_1})$.}
\label{F-4a}
\end{figure}

Define the matrix elements of $\pi(f)$ between semiclassical eigen-half-densities as the path
integral
\begin{equation}\label{me}
(\psi_{b_2}^{(2)}, \pi(f)\psi^{(1)}_{b_1})=\int \exp(\frac{i}{h}S[X,\eta]) f(X(\bullet)) DXD\eta
\end{equation}
Here the integral is taken over fields of the Poisson sigma model with the target space $\cP$
with boundary conditions $X|_{I_a}\in \cL_{a}\subset S\subset \cP$, $\eta|_{I_a}\in N^*_X\cL_a$, $a=1,2$
where $\cL_a=\cL_{b_a}^{(a)}$ are Lagrangian fibers of $\pi_{1,2}$. On the lower part of the boundary we have the same boundary condition with $\cL_a$ being replaced by $\Lambda$
and on the upper part of the boundary conditions are $\eta=0$ and $X$ is any.
These boundary conditions and the insertion of the function $f$ are illustrated on Fig. \ref{F-4a}.

Note that matrix elements (\ref{me}) do not depend on the choice of $\alpha$. In particular, for $M=T^*Q$ a combination of a formal diffeomorphism and a formal deformation of $\omega$  brings this star product to the one given by differential operators
(the one we used in the first section).

\begin{figure}[htb]
\includegraphics[height=4cm,width=8cm]{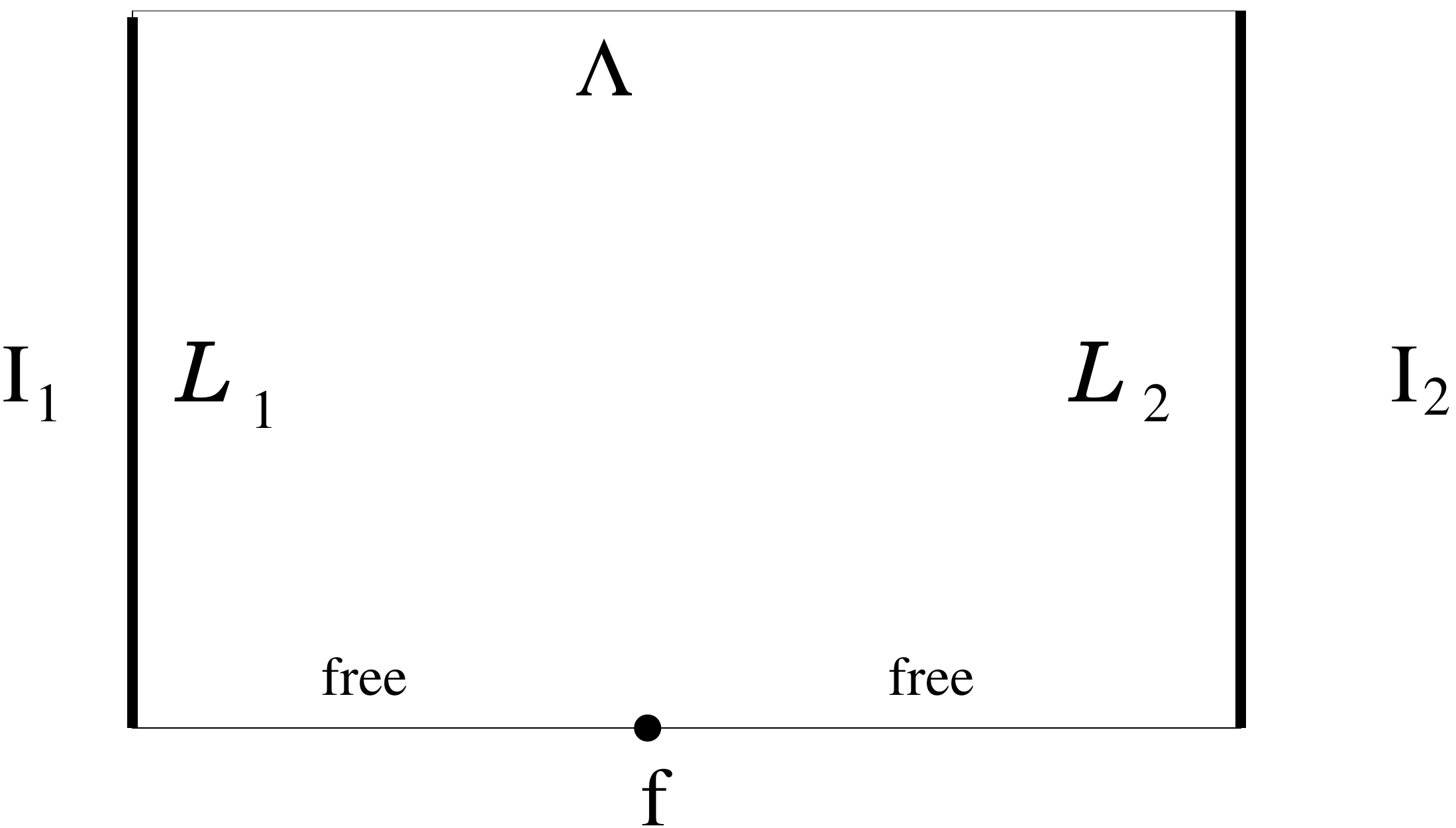}
\caption{The boundary conditions for the path integral representing matrix elements $(\psi_{b_1}^{(1)}, \pi(f)^*\psi^{(2)}_{b_2})=\overline{(\psi_{b_2}^{(2)}, \pi(f)\psi^{(1)}_{b_1})}$.}
\label{F-4b}
\end{figure}

Boundary conditions corresponding to the dual map $\pi(f)^*$ are shown on Fig. \ref{F-4b}.
It is natural to consider (\ref{me}) as the kernel of an integral operator $\pi_{21}(f): H_{P_1}^{1/2}\to H_{P_2}^{1/2}$.
In particular, when $P_1=P_2=P$, $\pi(F)$ is an endomorphism of $H_{P}^{1/2}$. But in this case the semiclassical
limit becomes quite singular. It is clear that the following identities hold:
\[
U_{P_3P_2}\pi_{21}(f)=\pi_{31}(f)=\pi_{32}(f)U_{P_2P_1}
\]
where $U_{P_2P_1}$ is the Blattner-Kostant-Sternberg kernel given by the scalar product which was 
the subject of previous sections. For the semiclassical asymptotic of (\ref{me}) we have
\begin{eqnarray}\label{pif}
\pi_{21}(f)(b_2,b_1)=\frac{C}{(2\pi h)^{n/2}}&\sum_{c\in \cL_{b_1}^{(1)}\cap \cL_{b_2}^{(2)}} e^{\frac{i}{h}S_{\gamma_1,\gamma_2}(c,b_1,b_2)+\frac{i\pi}{2}\mu_{\gamma_1,\gamma_2}(c)}\\ &\left|det\left(\frac{\pa^2 S_{\gamma_1,\gamma_2}(c,b_1,b_2)}{\pa b_1^i\pa b_2^j}\right)\right|^{1/2}f(c)(1+O(h))\sqrt{|db_1db_2|} \nonumber
\end{eqnarray}
where everything has the same meaning as in (\ref{sc-product}).

The path integral arguments used in the composition law of amplitudes also imply 
the following important property of matrix elements (\ref{me}). If $f$ is constant 
on fibers $\cL_{b_1}^{(1)}$ and $g$ is constant on fibers $\cL_{b_2}^{(2)}$ then
\[
\pi_{21}(f)(b_2,b_1)=U_{P_2P_1}(b_2,b_1)f(b_1) \ \ \pi_{21}(g)(b_2,b_1)=g(b_2)U_{P_2P_1}(b_2,b_1)
\]

Now we should check the homomorphism property:
\begin{equation}\label{comp}
\pi_{21}(f*g)(b_2,b_1)=\int_{B}\pi_{20}(f)(b_2,b)\pi_{01}(g)(b,b_1)
\end{equation}
where index $0$ corresponds to the Lagrangian fibration $P$ and $*$ is the Kontsevich's star product.

\begin{figure}[htb]
\includegraphics[height=4cm,width=8cm]{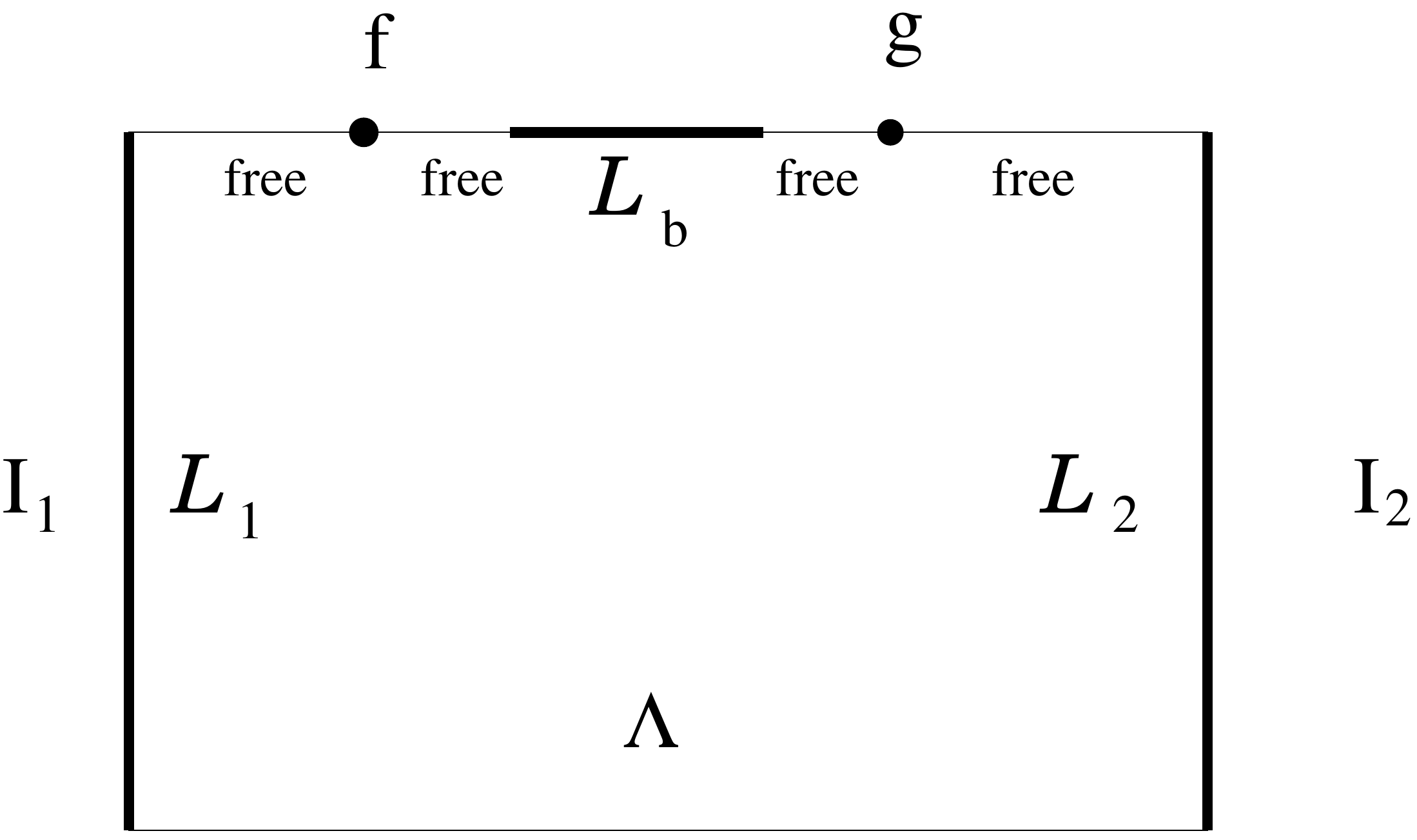}
\caption{Boundary conditions for the path integral interpretation of the integrand in (\ref{comp}).}
\label{F-5a}
\end{figure}

\begin{figure}[htb]
\includegraphics[height=4cm,width=8cm]{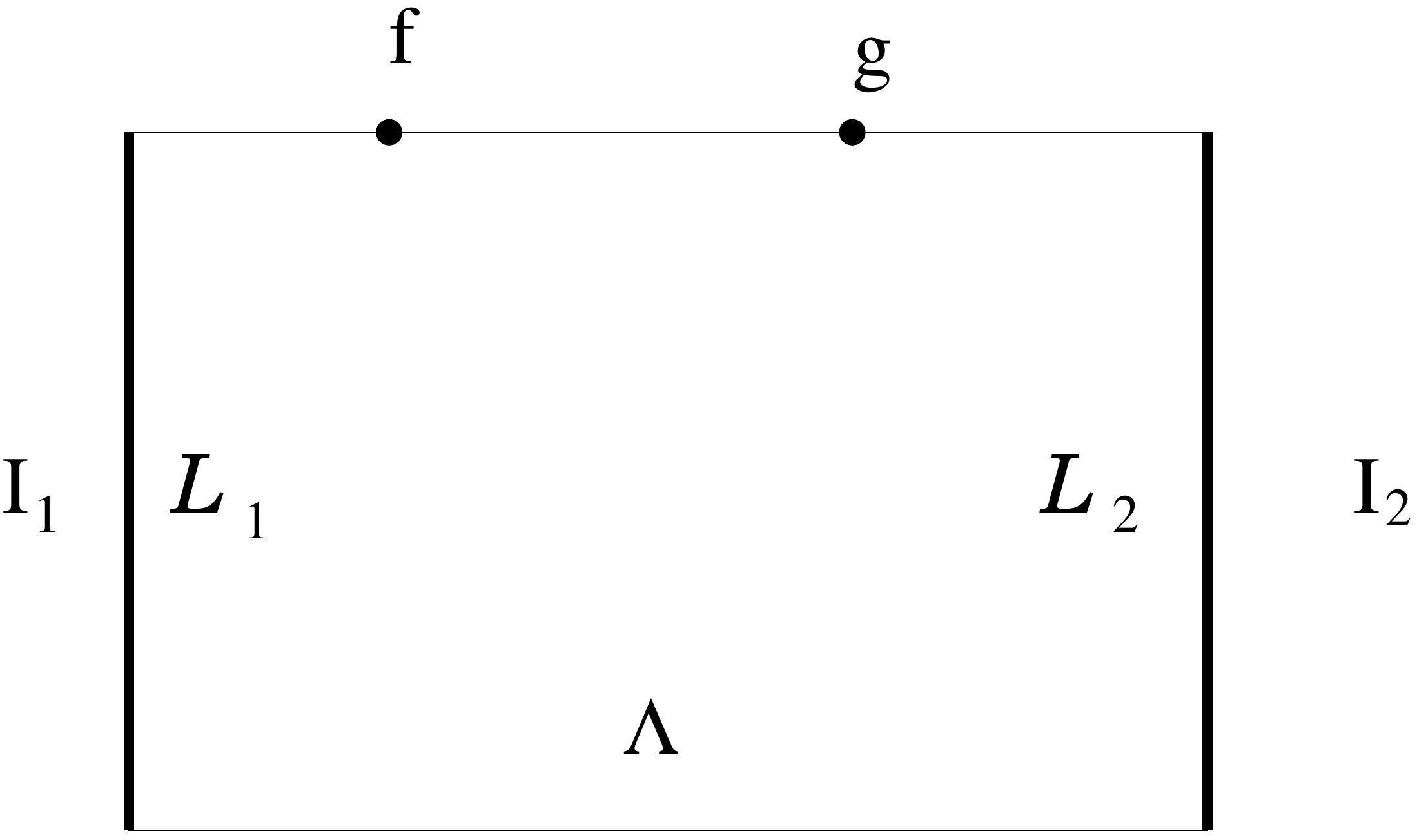}
\caption{Boundary conditions for the path integral interpretation  of the result of integration in the right side of (\ref{comp}).}
\label{F-5b}
\end{figure}

\begin{figure}[htb]
\includegraphics[height=5cm,width=8cm]{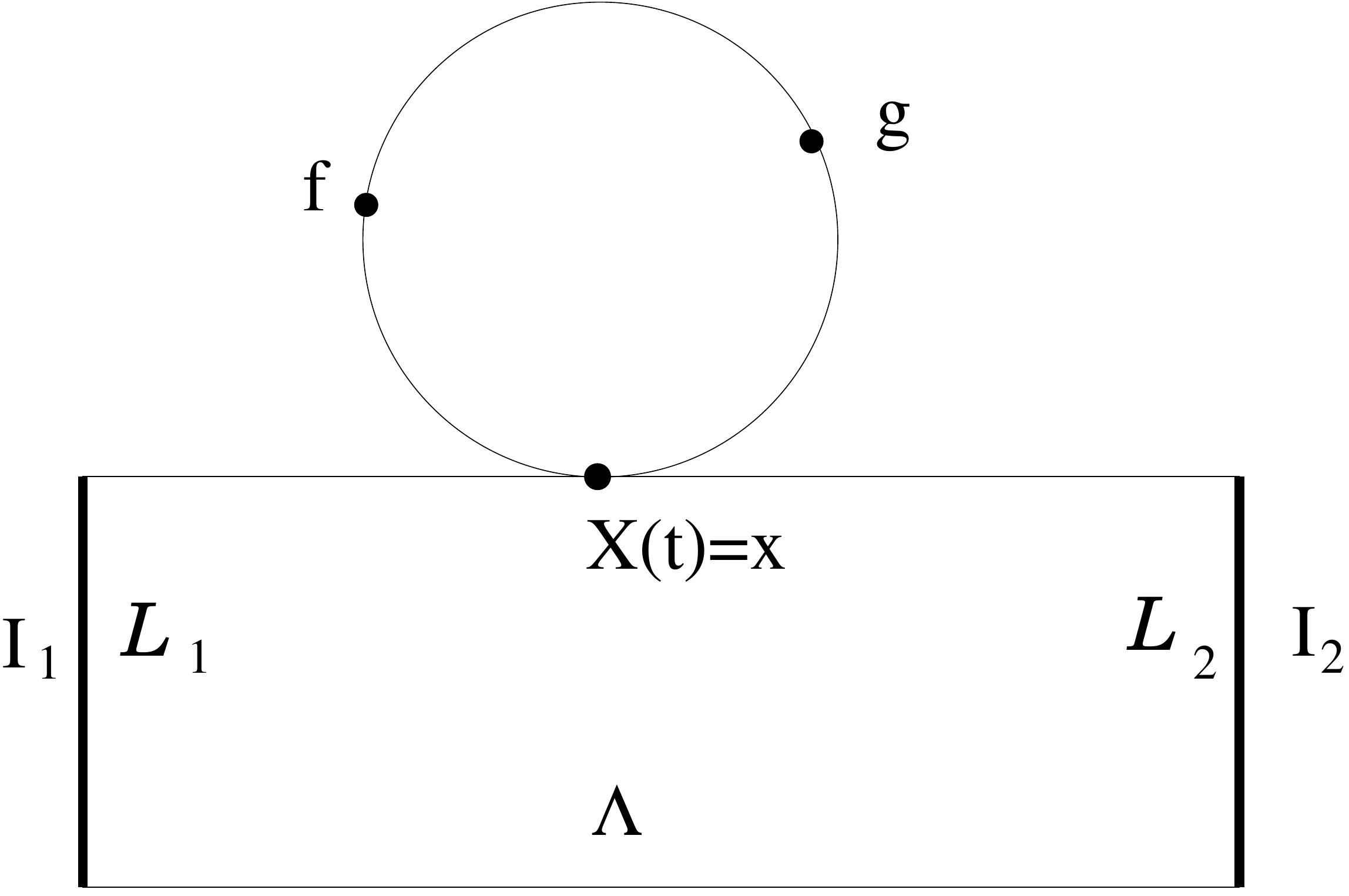}
\caption{Bounary conditions for the path integral interpretation of the left side of (\ref{comp}).}
\label{F-5c}
\end{figure}

This identity has the following path integral interpretation.
The partition function corresponding to Fig. \ref{F-5a} is equal to the partition function
corresponding to Fig. \ref{F-5b} and Fig. \ref{F-5c}. The arguments are the same as in case of the gluing properties of amplitudes.

According to the general philosophy of quantization of Poisson manifolds, symplectic leaves correspond to irreducible
representations\footnote{When $\cP$ is the dual space to a Lie algebra, this correspondence is at heart of
the orbit method, which assigns irreducible representations of a Lie group to co-adjoint orbits (provided necessary integrality conditions
are satisfied).
This is known as Kirillov-Kostant method.}. Thus, we expect that representations constructed in this way
are irreducible representations of $C_h(\cP)$.

\section{Conclusion} Throughout this paper we use real polarizations which are given by Lagrangian fibrations.
Such polarizations have clear semiclassical meaning in terms real symplectic geometry and analytical mechanics.

However such polarizations are typically quite singular. Complex polarizations  have richer
mathematical structures, are less singular,
allow tools from complex geometry, and in many important cases, from algebraic geometry. This is why
complex polarizations are more in use in mathematical literature on geometric quantization.
In physics literature, the corresponding representations of quantum observables are known
as the holomorphic realization of quantum mechanics. Topological quantum mechanics
in holomorphic representation is closely related the A-model in string theory \cite{W}
(see also \cite{B}).

Note that the path integral formula for (\ref{sc-product}) suggests that the torsion for the Poisson sigma model
on a disc with boundary conditions corresponding to TQM is given by the semiclassical Hessian in (\ref{sc-product}).

This article is only an outline. The detailed description of Feynman diagrams with proofs of basic statements  will be done 
in a separate publication.

\end{document}